\documentclass[twocolumn]{aastex631}

\usepackage{tabularx}
\usepackage{amssymb}

\begin{document}

\title{Phase-resolving the absorption signatures of water and carbon monoxide in the atmosphere of the ultra-hot Jupiter WASP-121b with GEMINI-S/IGRINS}

\author[0000-0003-3191-2486]{Joost P. Wardenier}
\altaffiliation{joost.wardenier@umontreal.ca}
\affiliation{Institut Trottier de Recherche sur les Exoplanètes, Université de Montréal, Montréal, Québec, H3T 1J4, Canada}
\affiliation{Atmospheric, Oceanic, and Planetary Physics, Clarendon Laboratory, University of Oxford, Oxford OX1 3PU, UK}

\author[0000-0001-9521-6258]{Vivien Parmentier}
\affiliation{Université Côte d’Azur, Observatoire de la Côte d’Azur, CNRS, Laboratoire Lagrange, France}

\author[0000-0002-2338-476X]{Michael R. Line}
\affiliation{School of Earth $\&$ Space Exploration, Arizona State University, Tempe AZ 85287, USA}

\author[0000-0003-4241-7413]{Megan Weiner Mansfield}
\affiliation{Steward Observatory, University of Arizona, Tucson, AZ 85719, USA}
\affiliation{NHFP Sagan Fellow}

\author[0000-0003-2278-6932]{Xianyu Tan}
\affiliation{Tsung-Dao Lee Institute, Shanghai Jiao Tong University, 520 Shengrong Road, Shanghai, People’s Republic of China}
\affiliation{School of Physics and Astronomy, Shanghai Jiao Tong University, 800 Dongchuan Road, Shanghai, People’s Republic of China}

\author[0000-0002-8163-4608]{Shang-Min Tsai}
\affiliation{Department of Earth Sciences, University of California, 900 University Ave, Riverside CA 92521, US}

\author[0000-0003-4733-6532]{Jacob L. Bean}
\affiliation{Department of Astronomy $\&$ Astrophysics, University of Chicago, Chicago, IL 60637, USA}

\author[0000-0002-4125-0140]{Jayne L. Birkby}
\affiliation{Astrophysics, University of Oxford, Denys Wilkinson Building, Keble Road, Oxford, OX1 3RH, UK}

\author[0000-0002-7704-0153]{Matteo Brogi}
\affiliation{Dipartimento di Fisica, Universit\`a degli Studi di Torino, via Pietro Giuria 1, I-10125, Torino, Italy}
\affiliation{INAF—Osservatorio Astrofisico di Torino, Via Osservatorio 20, I-10025 Pino Torinese, Italy}

\author[0000-0002-0875-8401]{Jean-Michel Désert}
\affiliation{Anton Pannekoek Institute for Astronomy, University of Amsterdam, Science Park 904, 1098 XH Amsterdam, The Netherlands}

\author[0000-0001-9552-3709]{Siddharth Gandhi}
\affiliation{Leiden Observatory, Leiden University, Postbus 9513, 2300 RA Leiden, The Netherlands}
\affiliation{Department of Physics, University of Warwick, Coventry CV4 7AL, UK}
\affiliation{Centre for Exoplanets and Habitability, University of Warwick, Gibbet Hill Road, Coventry CV4 7AL, UK}

\author[0000-0002-3052-7116]{Elspeth K. H. Lee}
\affiliation{Center for Space and Habitability, University of Bern, Gesellschaftsstrasse 6, CH-3012 Bern, Switzerland}

\author[0000-0002-3454-1695]{Colette I. Levens}
\affiliation{Atmospheric, Oceanic, and Planetary Physics, Clarendon Laboratory, University of Oxford, Oxford OX1 3PU, UK}

\author[0000-0002-1321-8856]{Lorenzo Pino}
\affiliation{INAF—Osservatorio Astrofisico di Arcetri, Largo Enrico Fermi 5, I-50125 Firenze, Italy}

\author[0000-0002-9946-5259]{Peter C. B. Smith}
\affiliation{School of Earth $\&$ Space Exploration, Arizona State University, Tempe AZ 85287, USA}



\begin{abstract}
Ultra-hot Jupiters are among the best targets for atmospheric characterization at high spectral resolution. Resolving their transmission spectra as a function of orbital phase offers a unique window into the 3D nature of these objects. In this work, we present three transits of the ultra-hot Jupiter WASP-121b observed with Gemini-S/IGRINS. For the first time, we measure the phase-dependent absorption signals of CO and H$_{\text{2}}$O in the atmosphere of an exoplanet, and we find that they are different. While the blueshift of CO increases during the transit, the absorption lines of H$_{\text{2}}$O become less blueshifted with phase, and even show a redshift in the second half of the transit. These measurements reveal the distinct spatial distributions of both molecules across the atmospheres of ultra-hot Jupiters. Also, we find that the H$_{\text{2}}$O signal is absent in the first quarter of the transit, potentially hinting at cloud formation on the evening terminator of WASP-121b. To further interpret the absorption trails of CO and H$_{\text{2}}$O, as well as the Doppler shifts of Fe previously measured with VLT/ESPRESSO, we compare the data to simulated transits of WASP-121b. To this end, we post-processes the outputs of global circulation models with a 3D Monte-Carlo radiative transfer code. Our analysis shows that the atmosphere of WASP-121b is subject to atmospheric drag, as previously suggested by small hotspot offsets inferred from phase-curve observations. Our study highlights the importance of phase-resolved spectroscopy in unravelling the complex atmospheric structure of ultra-hot Jupiters and sets the stage for further investigations into their chemistry and dynamics. 
\end{abstract}




\label{sec:intro}
\section{Introduction}

With an orbital period of 1.27 days, a bloated atmosphere, and an equilibrium temperature of \mbox{$\sim$2400 K}, WASP-121b (\citealt{Delrez2016}) is one of the best-studied ultra-hot Jupiters (UHJs) to date. A large number of observational campaigns, both from space and from the ground, have rendered WASP-121b a benchmark object when it comes to understanding the chemical composition, atmospheric dynamics, and 3D temperature structure of UHJs. Table \ref{tab:wasp121_parameters} provides an overview of some relevant parameters of the WASP-121 system.  


The first atmospheric studies of WASP-121b were performed with HST/WFC3, at low spectral resolution. \citet{Evans2016} reported water in the planet's transmission spectrum, while \citet{Evans2017} found emission features associated with water in its secondary-eclipse spectrum, indicating a thermal inversion on the dayside (e.g., \citealt{Hubeny2003,Fortney2005,Gandhi2019}). They also reported evidence for VO, an optical absorber potentially contributing towards this inversion, but its presence in emission was later ruled out by additional observations with the same instrument (\citealt{Mikal-Evans2019,Mikal-Evans2020}). Conversely, optical transmission spectra taken with HST/STIS and WFC3 (\citealt{Evans2018}) did lead to constraints on the VO abundance, {\color{black}{probably because these observations probe a different part of the atmosphere.}} The data also revealed strong excess absorption at near-UV wavelengths. At the same time, no evidence for TiO (another frequently postulated inversion-agent) was found, indicating that Ti-bearing species may be cold-trapped on the nightside of WASP-121b (e.g., \citealt{Spiegel2009,Parmentier2013,Hoeijmakers2022}).

In recent years, ground-based high-resolution spectroscopy {\color{black}{(HRS; $R\gtrsim$ 15,000)}} has painted a more detailed picture of the chemical inventory of WASP-121b. To date, HRS has led to the unambiguous detection of Na \textsc{i} (\citealt{Sindel2018}), Fe \textsc{i} (\citealt{Gibson2020}), H \textsc{i} (\citealt{Cabot2020}), Cr \textsc{i}, V \textsc{i}, Fe \textsc{ii} (\citealt{Ben-Yami2020}), Mg \textsc{i}, Ca \textsc{i}, Ni \textsc{i} (\citealt{Hoeijmakers2020}),  K \textsc{i}, Li \textsc{i}, Ca \textsc{ii} (\citealt{Borsa2021}), Sc \textsc{ii} (\citealt{Merritt2021}), Mn \textsc{i} (\citealt{Hoeijmakers2022}), Co \textsc{i}, Sr \textsc{ii}, Ba \textsc{ii} (\citealt{Azevedo2022}), CO, and H$_2$O (this work)\footnote{Of the studies listed here, only the observations by \citet{Hoeijmakers2022} were in emission. }. Notably, none of these studies detected {\color{black}{\mbox{Ti \textsc{i}}, TiO or VO}} in the transmission {\color{black}{or emission spectrum}} of WASP-121b (see also \citealt{Merritt2020}). While the non-detection of \mbox{Ti \textsc{i}} and TiO is consistent with the cold-trapping hypothesis, \citet{Hoeijmakers2020,Hoeijmakers2022} argued that their detection of V \textsc{i} should imply the presence of VO, but that their analyses were not sensitive enough to the latter. A more recent detection of V \textsc{i} and VO in the transmission spectrum of the canonical UHJ WASP-76b (\citealt{Pelletier2023}) supports the idea that these species exist together.

\begin{deluxetable}{ccccc}

{\color{black}{

    \tablewidth{0pt} 
    \tablecaption{A few parameters of the WASP-121 system. This table is an abridged version of Table 2 in \citet{Maguire2022}. References: D16 (\citealt{Delrez2016}); B20 (\citealt{Bourrier2020}); B21 (\citealt{Borsa2021}).}
    \tablehead{
        \multicolumn{1}{c}{Parameter} &
        \multicolumn{1}{c}{Value} &
        \multicolumn{1}{c}{Reference} &
    }
     \label{tab:wasp121_parameters}
      \startdata
           & \textbf{Stellar parameters} & \\
        $R_\star$ & 1.458 $\pm$ 0.030 R$_{\odot}$ & D16 \\
        $M_\star$ & 1.353$^{+0.080}_{-0.079}$ M$_{\odot}$ & D16 \\
        $T_{\rm eff}$ & 6460 $\pm$ 140 K & D16 \\
        Spectral type & F6V & D16 \\
        $K_\star$ & 0.177$^{+0.0085}_{-0.0081}$ km/s & B20 \\
        $V_{\rm sys}$ & 38.198 $\pm$ 0.002 km/s & B21 \\
        \hline
         & \textbf{Planetary parameters} & \\
        $P$ & 1.27492504$^{+1.5\cdot10^{-7}}_{-1.4\cdot10^{-7}}$ days & B20\\
$R_{\rm p}$ & 1.753 $\pm$ 0.036 R$_{\rm Jup}$& B20 \\
$M_{\rm p}$ & 1.157 $\pm$ 0.070 M$_{\rm Jup}$& B20 \\
$a_{\rm p}$ & 0.02596$^{+0.00043}_{-0.00063}$ AU & B20 \\ 
$i_{\rm p}$ & 88.49 $\pm$ 0.16$^{\circ}$ & B20 \\
$T_{\rm eq}$ & 2358 $\pm$ 52 K & D16 \\
$K_{\rm p}$ & $\sim$217.65 km/s & B21
    \enddata

}}
\vspace{-12pt}
\end{deluxetable}


\vspace{-12pt}

Phase-curve observations with TESS (\citealt{Bourrier2020b,Daylan2021}), HST/WFC3 (\citealt{Evans2022}), Spitzer/IRAC (\citealt{Morello2023}), and, most recently, JWST/NIRSpec (\citealt{Evans2023}) have yielded important constraints on the global temperature structure of WASP-121b, and the contrast between the dayside and the nigthside. While the (optical) TESS phase curves are consistent with a hotspot at the substellar point, the (infrared) HST and JWST phase curves show a peak just before secondary eclipse ($\sim$3$^\circ$ in phase for JWST), suggesting a small \emph{eastward} hotspot offset (assuming no clouds). Remarkably, the two Spitzer phase curves presented in \citet{Morello2023} indicate a \emph{westward} offset, possibly due to instrument systematics. An independent re-analysis of the same data by \citeauthor{Davenport2024} ({\color{xlinkcolor}{subm.}}) reports offsets that are consistent with zero and close to the JWST results (but less precise). At the same time, there is a growing body of literature claiming signs of temporal variability in the atmosphere of WASP-121b (\citealt{Wilson2021,Ouyang2023,Changeat2024}). {\color{black}{While weather cycles could play a role, we note that atmospheric variability on exoplanets is not well understood and largely remains an open question.}}

Another avenue that allows to explore the 3D structure and dynamics of UHJs is phase-resolving the Doppler shifts of their transmission spectra with HRS (e.g., \citealt{Wardenier2021,Wardenier2023,Savel2022,Beltz2023,Prinoth2023}). Because UHJs typically rotate by \mbox{$\sim$30$^\circ$} as they pass in front of their star (assuming that the orbit is roughly edge-on; \citealt{Wardenier2022}), a transit observation probes different regions of the atmosphere at different orbital phases. A notable observation in this regard is the Fe absorption signal of the UHJ WASP-76b (\citealt{Ehrenreich2020,Kesseli2021,Pelletier2023}), which becomes increasingly blueshifted during the first half of the transit. In the second half, its Doppler shift remains roughly constant. 3D modelling studies (\citealt{Wardenier2021,Savel2022}) demonstrated that this behaviour results from an asymmetry between the morning and evening terminator of the planet, such as a difference in scale height (i.e., temperature) or cloud cover. {\color{black}{Additionally, magnetic effects could play a role (\citealt{Beltz2023})}}. For WASP-121b, \citet{Bourrier2020} and \citet{Borsa2021} also reported a variation in the Doppler shift of Fe with orbital phase.

In \citet{Wardenier2023} we simulated the absorption signals of five chemical species, including Fe, CO, and H$_2$O, in a typical UHJ under various atmospheric conditions. We showed that species can exhibit different phase-dependent Doppler shifts, depending on their 3D distribution across the atmosphere. {\color{black}{This 3D distribution determines whether an atom or molecule absorbs}} light on the \emph{dayside} (refractories such as Fe, but also CO) or the \emph{nightside} of the planet (e.g., H$_2$O). Hence, observing the absorption signals of multiple species yields a more comprehensive picture of the 3D thermochemical structure and dynamics of an UHJ. 

In this paper, we perform the first phase-resolved measurements of the Doppler shifts of CO and H$_2$O in an UHJ, based on three transits of WASP-121b observed with \mbox{Gemini-S/}IGRINS. The structure of this work is as follows. Section \ref{sec:observations} summarises the observations and data reduction. In Section \ref{sec:cross-correlation}, we recover the absorption signals of CO and H$_2$O using cross-correlation analysis, and in Section \ref{sec:ccf_phase_dependence} we study their dependence on orbital phase. In Section \ref{sec:gcm_comparison}, we compare our absorption signals, as well as the Fe signal measured by \citet{Borsa2021}, to predictions from global circulation models (GCMs). Finally, Section \ref{sec:discussion} provides a discussion, followed by a summary and conclusion in Section \ref{sec:conclusion}. Appendix \ref{app:A} contains a few supplementary figures.

\section{Observations and data reduction} \label{sec:observations}

\subsection{Observations}

\begin{figure*}
\vspace{-40pt} 
\makebox[\textwidth][c]{\hspace{1pt}\includegraphics[width=1.2\textwidth]{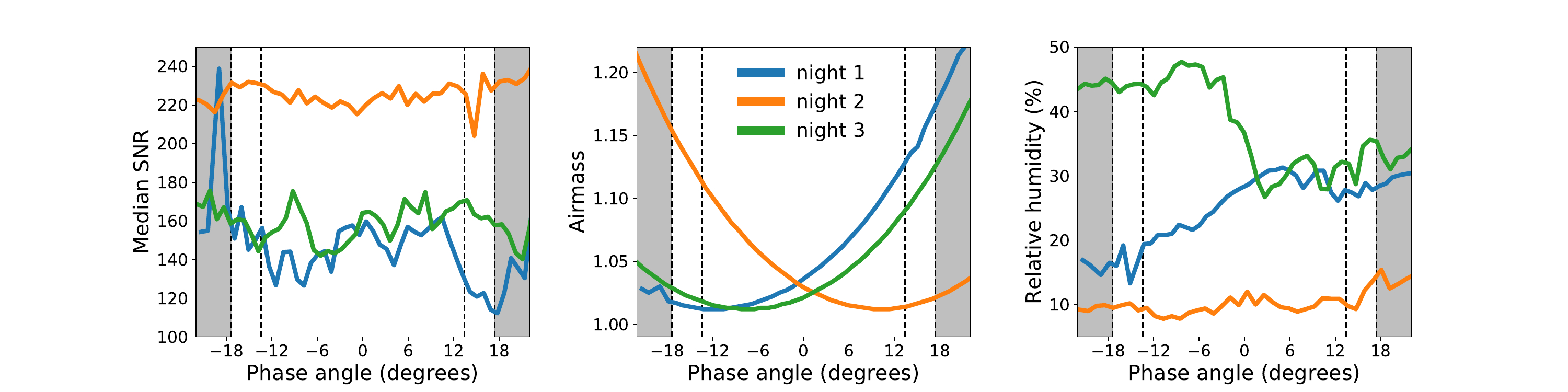}}

\vspace{-2pt}

\caption{The median signal-to-noise ratio over all (non-discarded) spectral orders (left panel), the airmass (middle panel), and the relative humidity (right panel) during each of the three observing nights as a function of the phase angle of WASP-121b. All values were taken directly from the headers of the \texttt{.fits} files output by the IGRINS pipeline package. In each plot, the dashed lines denote the ingress and egress phases of the transit, while the grey shaded regions mark the out-of-transit exposures.}
\vspace{5pt}
\label{fig:obs_diagnostics}
\end{figure*}

We observed three transits of WASP-121b with the \emph{Immersion GRrating INfrared Spectrometer} (IGRINS) on the 8.1-m Gemini-South Telescope, Cerro Pachón, Chile (\citealt{Yuk2010, Mace2018}). The observations were taken on 25 February, 10 December, and 24 December 2022 (dates in UTC) as part of programs GS-2022A-Q-242 and GS-2022B-Q-133 (PI: Wardenier). The observations are summarised in Table \ref{tab:transits}. Additionally, Fig. \ref{fig:obs_diagnostics} shows the signal-to-noise ratio (SNR), the airmass, and the relative humidity during each night as a function of the planet's orbital-phase angle.   

IGRINS has a spectral resolution $R\sim$ 45,000. The instrument comprises 54 spectral orders that cover a continuous wavelength range between 1.42 and 2.42 $\mu$m. To perform the observations, we used a standard ABBA nodding pattern and an exposure time of 100 s (nights 1 and 3) or 130 s (night 2) per frame. There is no particular reason for the longer exposure times during night 2 -- all observations were planned with a default exposure time of 100 seconds. However, 130 seconds is still well below limit at which Doppler smearing\footnote{Doppler smearing occurs when the radial velocity of the planet changes by more than one resolution element during a single exposure. For IGRINS, one resolution element amounts to $c/R \approx 6.7$ km/s, with $c$ the speed of light and $R$ the spectral resolution of the instrument. {\color{black}{At mid-transit, it takes $\sim$500 s for the radial velocity of WASP-12b to change by this value.}}} becomes an issue. On each night, we observed the full transit of WASP-121b, which takes just under 3 hours, as well as $\sim$30 minutes of stellar baseline before and after the transit. Including ingress and egress, the transit of WASP-121b occurs roughly between phase angles $\pm17^\circ$ (with $0^\circ$ corresponding to mid-transit), which is about a tenth of the planet's full orbit.

\vspace{-20pt}

\begin{deluxetable*}{cccccccccc}
    \tablewidth{0pt} 
    \tablecaption{Details of the transit observations of WASP-121b, performed with Gemini-S/IGRINS.}
    \tablehead{
        \multicolumn{1}{c}{} &
        \multicolumn{1}{c}{Date in UTC} &
        \multicolumn{1}{c}{Time window} &
        \multicolumn{1}{c}{Mid-transit} &
        \multicolumn{1}{c}{Number of} &
        \multicolumn{1}{c}{Exposure} &
        \multicolumn{1}{c}{Typical} &
        \multicolumn{1}{c}{Average} &
        \multicolumn{1}{c}{Average} &
        \\
        \multicolumn{1}{c}{} & 
        \multicolumn{1}{c}{(dd/mm/yyyy)} &
        \multicolumn{1}{c}{in UTC} &
         \multicolumn{1}{c}{in UTC} &
        \multicolumn{1}{c}{AB pairs} &
        \multicolumn{1}{c}{time per frame} &
        \multicolumn{1}{c}{SNR} &
        \multicolumn{1}{c}{airmass} &
        \multicolumn{1}{c}{humidity} &
    }
    \startdata
        \textbf{Night 1} & 25/02/2022 & 00:45--04:39 & 02:33 & 50 & 100 s & 147 & 1.06 & 25$\%$ \\
        \textbf{Night 2} & 10/12/2022 & 03:33--07:50 & 05:45 & 45 & 130 s & 225 & 1.08 & 10$\%$ \\
        \textbf{Night 3} & 24/12/2022 & 04:16--08:27 & 06:20 & 53 & 100 s & 163 & 1.04 & 39$\%$ \\
    \enddata
    \tablecomments{The typical SNR was computed by averaging med($\vec{s}$) over all in-transit exposures. Here, med() is the median operator, while $\vec{s}$ is a vector containing the median SNR values of all spectral orders used in the data analysis.}
    
\label{tab:transits}
\end{deluxetable*}

\subsection{Data reduction}
\label{subsec:blaze_correction}

An initial data reduction was carried out with the IGRINS pipeline package\footnote{Available from \href{https://github.com/igrins/plp}{ github.com/igrins/plp}.} (PLP, \citealt{Sim2014,Lee_PIP_2017}) by the instrument team. We perform further wavelength calibration and a barycentric-velocity correction using routines from the \texttt{IGRINS\_transit} package\footnote{Available from \href{https://github.com/meganmansfield/IGRINS_transit}{github.com/meganmansfield/IGRINS$\_$transit}.} (\citealt{igrins_transit}), which builds on earlier work from \citet{Line2021}. For each observation, \texttt{IGRINS\_transit} produces a data cube that contains normalised fluxes as a function of wavelength, spectral order, and time (i.e., orbital phase). After wavelength cropping, each order covers 1848 pixels. We discard orders 1, 23$-$27, 53, and 54, which are situated near the edges of the instrument filters, and which coincide with the strongest telluric absorption lines. This results in SNR values that are much lower compared to other parts of the spectrum.

Next, {\color{black}{to correct for throughput variations over time}}, we put all spectra on a ``common blaze'' (e.g., \citealt{Gibson2020}) by dividing the fluxes at each orbital phase by a second-order polynomial (we do this for each order separately). This polynomial is a fit to the fluxes in each pixel divided by their median over time. We also perform a baseline subtraction. To this end, we convolve the fluxes with a Gaussian kernel (FWHM = 80 pixels) and we subtract the result from the original spectrum. This removes broadband features from the data that could impact the cross-correlation analysis. Finally, we cut off 100 pixels at the edges of each order to get rid of boundary effects. Because these pixels have inherently lower SNR values (due to the shape of the blaze function), the loss in planet signal is at most a few percent. At the end, each order comprises 1648 pixels.

\subsection{Correcting for telluric absorption}
\label{subsec:telluric_corr}

The fluxes contained in the data cubes are a combination of the stellar spectrum, telluric absorption (especially in the infrared), the planet signal, and photon noise. When it comes to extracting the planet signal from the data, HRS leverages the fact that stellar and telluric lines remain effectively \emph{stationary} during the observation, while the spectrum of the planet undergoes a Doppler shift induced by its orbital motion (e.g., \citealt{Snellen2010,Brogi2012,Birkby2018}).

To separate the planet signal from the systematics, we perform a singular value decomposition (SVD) on each spectral order, for each night (e.g., \citealt{DeKok2013,Giacobbe2021}). The sum of the first $n$ SVD components (which are all rank-1 matrices) is the best rank-$n$ approximation of the data. Therefore, the components associated with the largest singular values will capture the strong, \mbox{(quasi-)}stationary lines caused by telluric/stellar absorption, while the residuals will contain the much weaker planet spectrum buried in noise.

There is no golden rule prescribing how many SVD components to remove from the data -- removing too few will cause systematics to persist, while removing too many will affect the planet signal. Moreover, recent work by \citet{Smith2023} on IGRINS data shows that it may be advantageous to remove a \emph{different} number of SVD components per order and per night, depending on the observing conditions (humidity, airmass, etc.) and the degree of telluric absorption within an order\footnote{For further discussion, see \citet{Nugroho2017}, \citet{Spring2022}, \citet{Cheverall2023}, and \citet{Klein2024}.}. However, designing such an optimisation process is non-trivial, and HRS studies generally opt to remove the same number of SVD components from each order for a given observing night (e.g., \citealt{Line2021,vanSluijs2022,Boucher2023,Brogi2023,Pelletier2023,Smith2023}). In this work, we adopt a similar approach. Yet, to remain as agnostic as possible with regards to the telluric correction, we carry out separate analyses for $3 \leq n \leq 18$, with $n$ the \emph{fixed} number of SVD components removed from each of the orders, for each night. The motivation for the lower limit of $n$ = 3 is that, from this value onwards, the $K_{\text{p}}$--$V_{\text{sys}}$ maps of the combined transits (see Fig. \ref{fig:kp_vsys_maps}) show clear detections of CO and H$_{\text{2}}$O near the expected planet position. {\color{black}{The upper limit $n$ = 18 was chosen because this is where the SNR associated with the H$_{\text{2}}$O detection first drops below 4 (Fig. \ref{fig:srn_svd_components}).}} 

Additionally, it should be noted that for each of the three observing nights, there is a partial overlap between the telluric absorption lines and the spectrum of WASP-121b  in velocity space (see Fig. \ref{fig:ccf_maps}). That is, $V_{\text{bary}} \approx V_{\text{planet}}(\phi_i)$ for some phase angle $\phi_i$, with $V_{\text{bary}}$ the barycentric velocity and $V_{\text{planet}}$ the planet velocity\footnote{In each of our observations, the barycentric velocity changes by less than 1 km/s (see Fig. \ref{fig:ccf_maps}). Conversely, the planet velocity changes by more than 120 km/s during the transit.}. Such an overlap is nearly unavoidable in transit observations of short-period planets like WASP-121b, as their radial velocity takes on a wide range of values centered around ($V_{\text{bary}} + V_{\text{sys}}$), with $V_{\text{sys}}$ the systemic velocity of the planet. In this light, it is especially important to perform the data analysis for different values of $n$, as choosing a single number of SVD components could lead to remaining systematics being interpreted as planetary signal. 

\section{Cross-correlation analysis} \label{sec:cross-correlation}

\subsection{CCF maps} \label{sec:ccf_maps}

Once the $n$ largest SVD components are subtracted from the data, we are left with residual fluxes containing the planet spectrum buried in noise. To recover the signals of individual molecules in the atmosphere of WASP-121b, we cross-correlate the residual fluxes with a template spectrum at each orbital phase. This gives rise to a 2D cross-correlation function (CCF):

\begin{equation} \label{eq:def_cross_correlation}
\text{CCF}(\phi, v) = \sum_{i=1}^{N_{\text{orders}}} \sum_{j=1}^{N_{\text{pixels}}} x_{i,j}(\phi) \ T_{i,j}(v). 
\end{equation}

\noindent In this equation, $x_{i,j}(\phi)$ is the residual flux in the $j$-th pixel of the $i$-th order at phase angle $\phi$. $T_{i,j}(v)$ is the value of the template spectrum (with units of transit depth) shifted by a velocity $v$, evaluated at the wavelength corresponding to $x_{i,j}$. 

Our template spectra are based on the SPARC/MITgcm model of WASP-121b from \citet{Parmentier2018}. {\color{black}{The chemical abundances of this model were computed assuming chemical equilibrium and a solar metallicity (the left column in Fig. \ref{fig:equatorial_maps} shows the equatorial plane of the GCM, including the abundances of CO, H$_2$O, and OH)}}. To obtain the template spectrum for a chemical species $X$, we post-process the model with gCMCRT (\citealt{Lee2022}), a 3D Monte-Carlo raditive transfer code. For each template, we only include the opacities of species $X$ and those of the continuum\footnote{We include continuum opacities of H$_2$, He, and H scattering, collision-induced absorption (CIA) by H$_2$-H$_2$ and H$_2$-He, and bound-free and free-free transitions associated with H$^-$. For references to the opacity data, see Table 2 in \citet{Lee2022b}.}. We also account for Doppler-broadening of absorption lines caused by planet rotation. As in \citet{Wardenier2023}, all templates are computed at \emph{mid-transit}, at $R\sim$ 135,000. With regards to Eq. \ref{eq:def_cross_correlation}, this means that the template is only a function of velocity shift, but not of orbital phase. Once the (rotationally broadened) spectra are obtained, we convolve them with a Gaussian kernel whose FWHM corresponds to the spectral resolution of the instrument ($R\sim$ 45,000). Prior to evaluating the CCF, we subtract the continuum from the templates such that they have a horizontal baseline. The cross-correlation is performed for $v$ between $\pm$300 km/s, with steps of 0.5 km/s.

\begin{figure*}
\vspace{-50pt} 
\makebox[\textwidth][c]{\hspace{5pt}\includegraphics[width=1.0\textwidth]{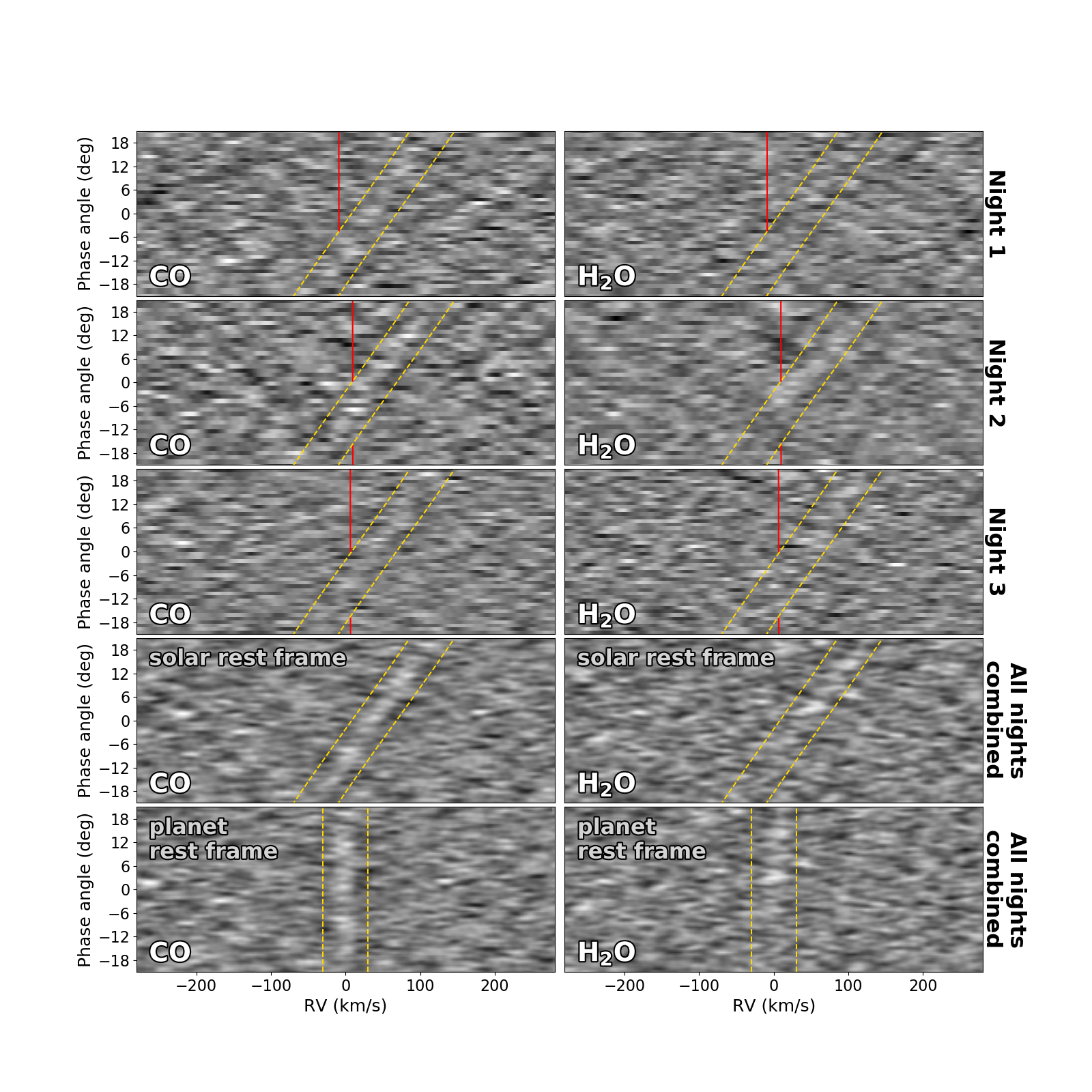}}

\vspace{-35pt}

\caption{Cross-correlation (CCF) maps showing the CO and H$_{\text{2}}$O signals of WASP-121b, obtained when removing 5 SVD components from the data. Lighter colours imply a more positive correlation with the template {\color{black}{(see Section \ref{sec:ccf_maps})}}. The first three rows depict the absorption signal observed during the individual nights in the solar rest frame (i.e., corrected for barycentric motion). The yellow dashed lines mark the expected planet trail $\pm$30 km/s, and the red lines show the barycentric velocity during each night. The fourth row shows the sum of the three absorption trails. The bottom row contains the same CCF map, but shifted into the \emph{planet rest frame}, where the signal manifests as a vertical trail at $v$ = 0 km/s.}

\vspace{5pt}

\label{fig:ccf_maps}
\end{figure*}

To further assess the robustness of our analysis, we also perform a test in which we apply the SVD to the data cube from \texttt{IGRINS$\_$transit} directly (i.e., without performing the blaze correction and baseline subtraction described in Section \ref{subsec:blaze_correction}). However, before computing the CCF, we first inject the template spectrum into the main SVD components and then ``recover'' it through a second SVD, as is routinely done in retrievals on high-resolution datasets (e.g., \citealt{Brogi2019,Line2021,Brogi2023,Smith2023}). This gives rise to a new template $\tilde{T}_n$, with $n$ denoting the number of removed SVD components. $\tilde{T}_n$ is a more realistic representation of the planet signal, as it accounts for the subtle changes that may have been imparted on the spectrum during the telluric correction. Reassuringly, as shown in Fig. \ref{fig:appendix_doppler_comparison} in Appendix \ref{app:A}, the phase-dependence of the Doppler shifts that we observe for CO and H$_{\text{2}}$O does not change much with this alternative approach.

\begin{figure*}
\vspace{-20pt} 
\makebox[\textwidth][c]{\hspace{5pt}\includegraphics[width=1.28\textwidth]{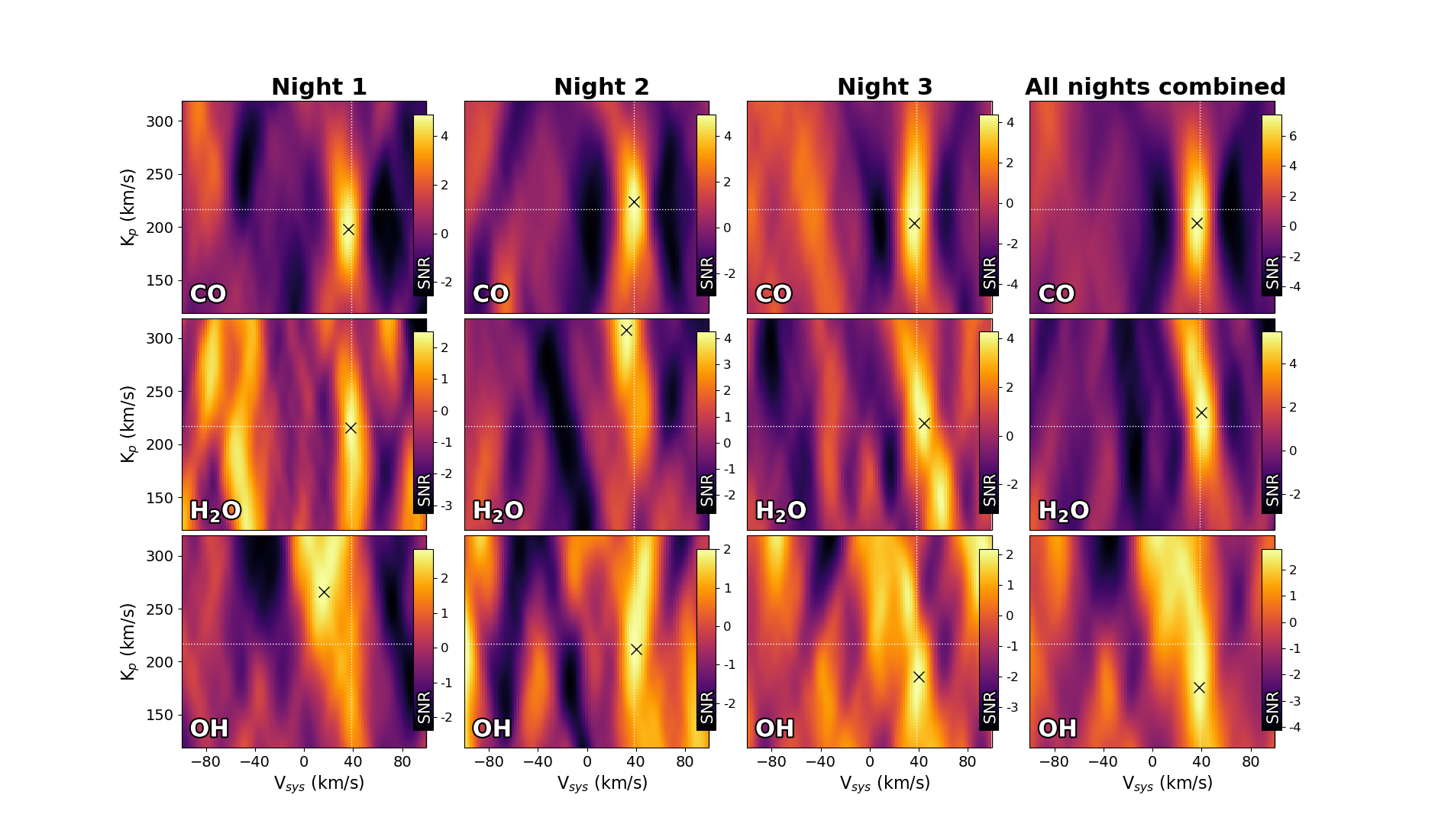}}

\vspace{-20pt}

\caption{$K_{\text{p}}$--$V_{\text{sys}}$ maps for CO, H$_{\text{2}}$O, and OH when removing 5 SVD components from the data. The first three columns are obtained from the individual transits, while the right column shows the maps when combining the three observations. The white dotted lines mark the known ($K_{\text{p}}$ $\sim$ 218 km/s, $V_{\text{sys}}$ $\sim$ 38 km/s) values of WASP-121b (e.g., \citealt{Borsa2021}), with the black crosses indicating the location associated with the maximum SNR value in the shown domain. The SNR values were computed using the $\sigma$-clipping method (see text and Fig. \ref{fig:srn_svd_components}), which produces slightly more conservative detections for CO and H$_{\text{2}}$O. The SNR values for OH are not significant enough to claim a detection, but the maps are included here for reference.}

\vspace{3pt}

\label{fig:kp_vsys_maps}
\end{figure*}

The CCF maps of CO and H$_{\text{2}}$O are shown in the first three rows of Fig. \ref{fig:ccf_maps}, which each row corresponding to one observing night. Besides CO and H$_{\text{2}}$O, we also searched for the presence of OH, TiO, VO, HCN and H$_{\text{2}}$S, but we did not obtain significant detections\footnote{References to line lists: CO (\citealt{Li2015}), H$_2$O (\citealt{Polyansky2018}), OH (\citealt{Rothman2010}), TiO (\citealt{Mckemmish2019}), VO (\citealt{McKemmish2016}), HCN (\citealt{Barber2014}) and H$_2$S (\citealt{Azzam2016}).}. We find very tentative evidence for OH (see Section \ref{sec:kp_vsys} and Fig. \ref{fig:kp_vsys_maps}), but the signal is not strong enough to claim a detection. 


As demonstrated in Fig. \ref{fig:ccf_maps}, the planetary trails in the CCF maps are relatively weak. Also, the planet signal on night 2 seems to be strongest when it coincides with the barycentric velocity, which warrants caution. This makes it challenging to assess the phase-dependent behaviour of CO and H$_{\text{2}}$O on a night-by-night basis. Thus, in order to enhance the signal and average any non-planetary contributions over all nights, we combine the three CCF maps, both in the solar rest frame (fourth row in Fig. \ref{fig:ccf_maps}) and the planet rest frame (bottom row). In the solar rest frame, the planet is visible as a diagonal trail. In the planet rest frame, the planet signal manifests as a vertical trail centered at $v = 0$ km/s (modulo any anomalous Doppler shifts). To transform the data to the planet rest frame we use $K_{\text{p}}$ = 217.65 km/s and \mbox{$V_{\text{sys}}$ = 38.20 km/s}, which are the same values as those used in \citet{Borsa2021}\footnote{F. Borsa, private communication}. Before adding the CCF maps together, we interpolate them onto a common grid with 1,000 points along the phase axis. Upon visual inspection, there appear to be no strong telluric artefacts {\color{black}{(i.e., clearly structured features outside the trail)}} in the combined CCF maps.    


\label{sec:kp_vsys}
\subsection{$K_{\text{p}}$--$V_{\text{sys}}$ maps}

Fig. \ref{fig:kp_vsys_maps} shows the $K_{\text{p}}$--$V_{\text{sys}}$ maps for CO, H$_{\text{2}}$O, and OH obtained when removing $5$ SVD components from the data. As in \citet{Wardenier2023}, we compute the $K_{\text{p}}$--$V_{\text{sys}}$ maps by integrating the values in the CCF map along ``orbital trails'' of the form $v(\phi) = V_{\text{sys}} + K_{\text{p}} \sin(\phi)$. This gives

\begin{equation} \label{eq:kpvsys_sum}
    \text{SNR}(K_{\text{p}}, V_{\text{sys}}) = \frac{1}{a} \sum_i^{N_{\text{phases}}} \text{CCF} \Big(\phi_i, v(\phi_i) \Big) - b,
\end{equation}

\noindent with SNR the signal-to-noise ratio at point $(K_{\text{p}},V_{\text{sys}})$. Furthermore, {\color{black}{$a$ is a scaling factor}}, $b$ is a baseline value such that the median of the map is zero, and $N_{\text{phases}}$ is the number of points along the phase axis of the CCF map. We integrate the signal between $\phi = \pm$17 degrees, which is the phase range covered by the transit of WASP-121b. For each orbital phase, we compute the CCF at $v(\phi_i)$ by performing a linear interpolation between the CCF values at the two closest radial velocities.

\begin{figure}

\vspace{3pt}

{\hspace{-10pt} \includegraphics[width=0.5\textwidth]{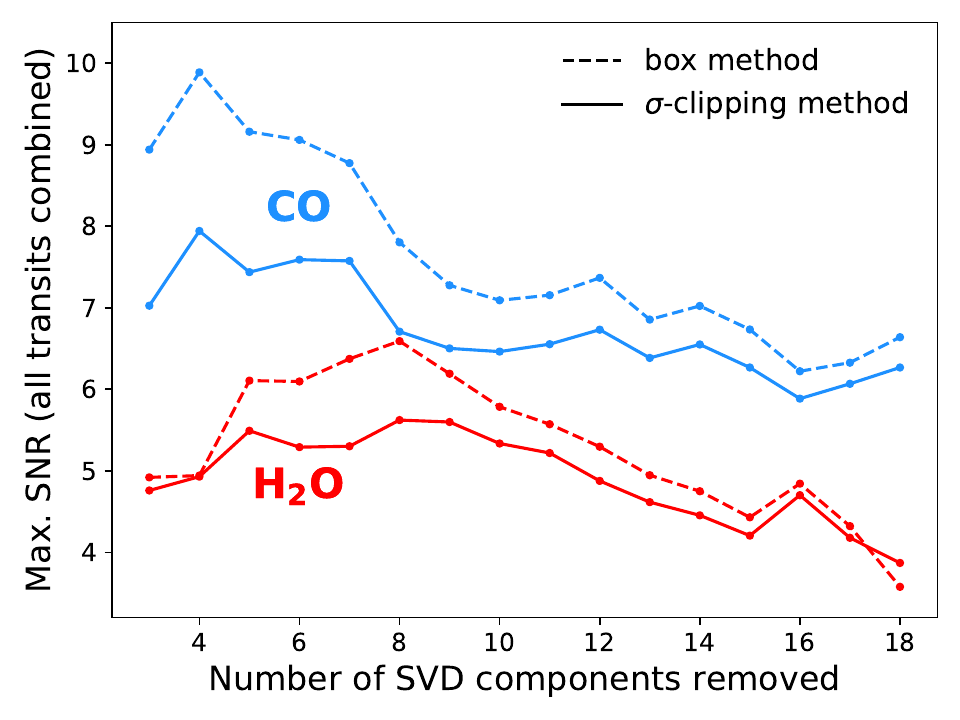}}

\caption{SNR associated with the CO and H$_{\text{2}}$O detections in the atmosphere of WASP-121b when combining the three transits, as a function of the number of SVD components removed from the data. The SNR is computed using two methods (see text). Overall, the $\sigma$-clipping method produces more conservative values for the SNR than the box method.}
\label{fig:srn_svd_components}
\end{figure}

We calculate the $K_{\text{p}}$--$V_{\text{sys}}$ maps for $K_{\text{p}}$ between $\pm$350 km/s and $V_{\text{sys}}$ between $\pm$150 km/s. Fig. \ref{fig:kp_vsys_maps} only shows a subregion of this domain, zooming in on the detection peaks (see Fig. \ref{fig:appendix_kpvsys} in Appendix \ref{app:A} for plots across the whole domain). We use two methods to determine the value of $a$ in Eq. \ref{eq:kpvsys_sum}. The ``box method'' (e.g., \citealt{Line2021,Nortmann2024}) sets $a$ equal to the standard deviation of all SNR values in a region (a box) far away from the expected planet position. In this work, we use the region $K_{\text{p}} < 0$ km/s. The ``$\sigma$-clipping method'' (e.g., \citealt{Kasper2021,WeinerMansfield2024}) takes the standard deviation of the whole map, but iteratively rejects values that deviate by more than 3$\sigma$ from the median (we use 4 iterations). The $K_{\text{p}}$--$V_{\text{sys}}$ maps in Fig. \ref{fig:kp_vsys_maps} show the SNR values obtained with the $\sigma$-clipping method.






In each of the individual nights, we detect CO at \mbox{SNR $\gtrsim$ 4}. When combining the observations\footnote{When combining the $K_{\text{p}}$--$V_{\text{sys}}$ maps, we first add the ``non-normalised'' maps together (i.e., with $a$ = 1 and $b$ = 0), and then compute the values of $a$ and $b$ for the sum (see Eq. \ref{eq:kpvsys_sum}).}, the signal is boosted to SNR $\gtrsim$ 7. Detecting H$_{\text{2}}$O based on an individual transit appears to be much harder. On night 1, there is a peak at the expected planet position, but its value is not significant. On night 2, we find a relatively strong H$_{\text{2}}$O peak, but it is offset by $\sim$100 km/s along the $K_{\text{p}}$ axis (possibly due to telluric effects). On night 3, there is a peak with SNR $\sim$ 4 near the expected planet position, but we find a similarly strong feature at a lower $K_{\text{p}}$ value. It is only when combining the observations that an unambiguous signal (SNR $\gtrsim$ 5) emerges near the known $(K_{\text{p}},V_{\text{sys}})$ of the planet. When it comes to OH, we find hints of its presence in the atmosphere of WASP-121b, but the SNR values in Fig. \ref{fig:kp_vsys_maps} are not significant. We will briefly reflect on this non-detection in Section \ref{subsec:non-detection}.


\begin{figure}

\vspace{-16pt}

{\hspace{-20pt} \includegraphics[width=0.565\textwidth]{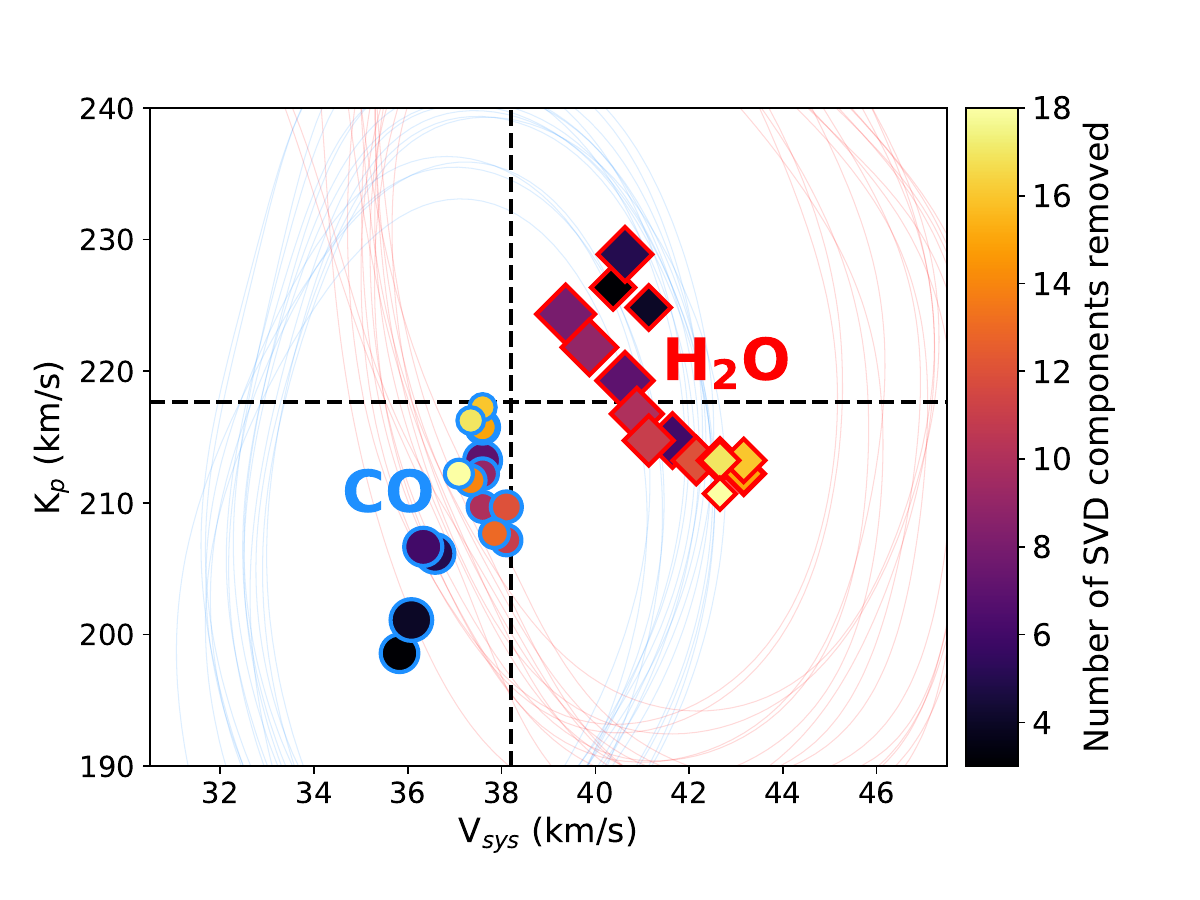}}

\vspace{-6pt}

\caption{The location of the detection peaks of CO (circles with blue edges) and H$_{\text{2}}$O (diamonds with red edges) in the combined $K_{\text{p}}$--$V_{\text{sys}}$ map when different numbers of SVD components are removed from the data (indicated by the colour scale). The greater the size of a symbol, the higher the associated detection SNR. The dashed lines mark the known ($K_{\text{p}}$, $V_{\text{sys}}$) values of the planet. The coloured contours in the background denote SNR = max(SNR) $-$ 1 for each of the points.}
\label{fig:maxima_positions_svd}
\end{figure}

Fig. \ref{fig:srn_svd_components} shows the detection SNR obtained as a function of the number of SVD components removed from the data. Depending on the method used, we find a maximum SNR of 8--10 for CO (when removing 4 SVD components from all orders for all nights) and a maximum SNR of 5.5--6.5 for H$_{\text{2}}$O (when removing 8 components). Furthermore, Fig. \ref{fig:maxima_positions_svd} shows the ($K_{\text{p}}$, $V_{\text{sys}}$) values at which the detection peaks occur for a given number of removed SVD components.

\begin{figure*}
    \centering

    \vspace{-15pt} 
    
    \begin{minipage}[c]{\textwidth}
    \makebox[\textwidth][c]{
    \includegraphics[width=1.2\textwidth]{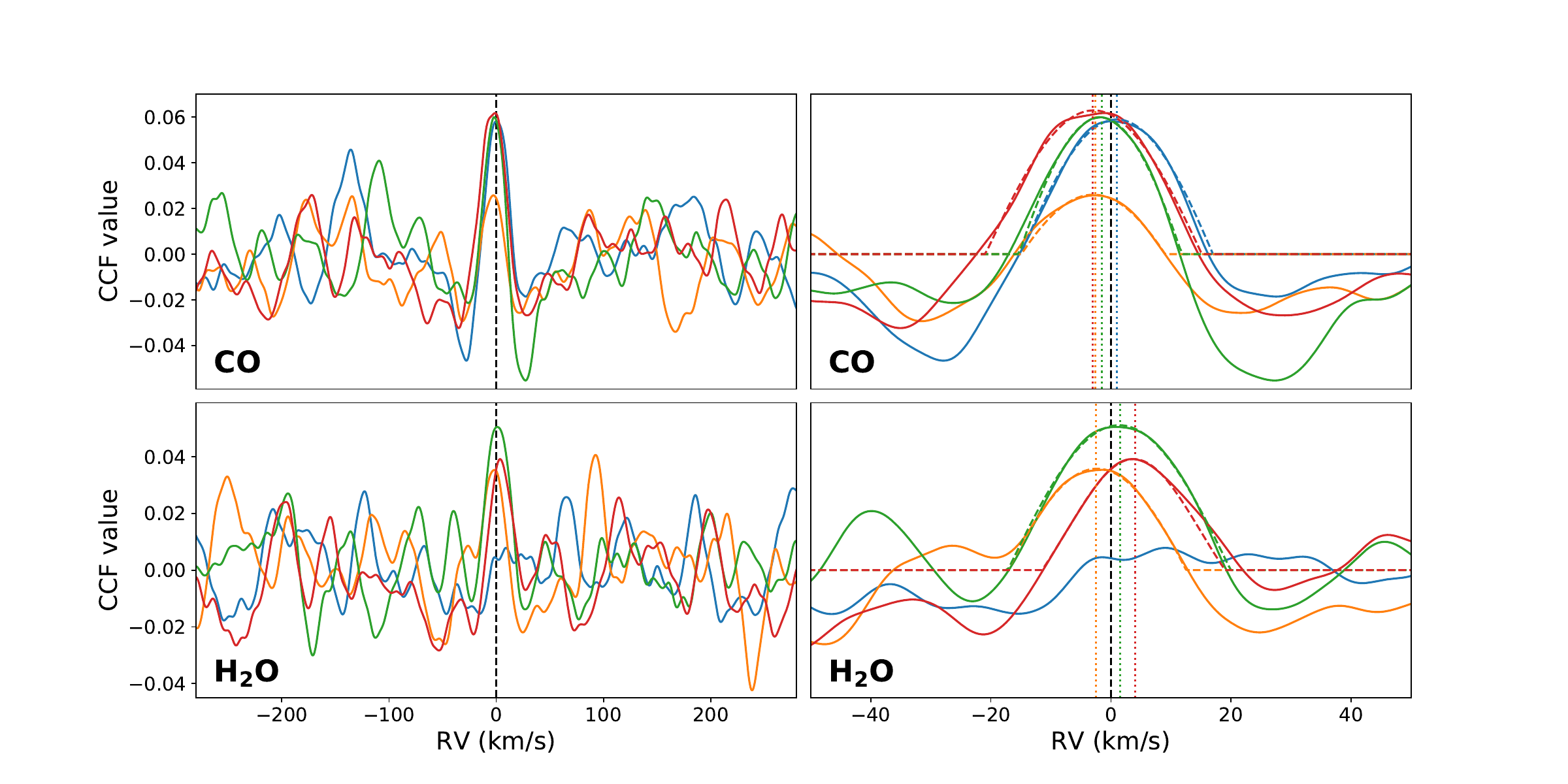}}
    \end{minipage}

     \vspace{-299pt} 

    \begin{minipage}[c]{\textwidth}
    \makebox[\textwidth][c]{
    \includegraphics[width=1.1\textwidth]{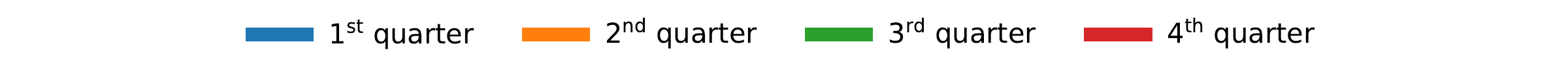}}
    \end{minipage} 

    \vspace{238pt} 

    \begin{minipage}[c]{1\textwidth}
      \vspace{25pt}
	  \caption{1D CCF signals associated with CO (top row) and H$_{\text{2}}$O (bottom row) in the planet rest frame, obtained when removing 5 SVD components from the data. The 1D CCFs were computed by splitting the CCF map into four quarters and summing the values along the phase axis. Each quarter spans $\sim$7 degrees in phase. The left panels show the 1D CCFs across a wide range of radial velocities, while the right panels show zoomed-in versions of the same plots. {\color{black}{Furthermore, the dashed curves show Gaussians that were fit to the peaks of the 1D CCFs to measure the Doppler shift of the planet spectrum as a function of orbital phase (see Fig. \ref{fig:doppler_svd}). The vertical dotted lines mark the peak positions of the Gaussian fits.}}} 
   \label{fig:1d_ccfs}
    \end{minipage}
\end{figure*}

{\color{black}{Finally, with regards to the H$_{\text{2}}$O signal, it should be noted that the $\sim$100 km/s $K_{\text{p}}$ offset on night 2 cannot be physical. This is likely due to some remaining systematics in the data. Therefore, by itself, the H$_{\text{2}}$O signal from night 2 should be treated with caution. However, we still include night 2 in our combined H$_{\text{2}}$O observation for several reasons. Firstly, the observation from night 2 still contains valid planet information, as demonstrated by the strong CO detection (for which the peak is not shifted) on the same night. Also, the $K_{\text{p}}$--$V_{\text{sys}}$ map of H$_{\text{2}}$O does feature a secondary (but less prominent) peak closer near the expected planet position. Furthermore, the observations from night 2 suffered from the lowest humidity and showed the highest SNR (see Table \ref{tab:transits} and Fig. \ref{fig:obs_diagnostics}), so in principle these data should contain the highest-quality signal. In the end, the fact that we combine three independent transits makes the combined observation more robust against noise effects from the individual nights. This is because the complementary planet signals add up, while the impacts of noise and systematics are ``diluted'' as they are averaged over all nights. Given the currently available data, this is the best we can do to maximise the H$_{\text{2}}$O signal. Additional transit observations of WASP-121b will be needed to confirm our measurements.}}

\section{The phase-dependence of the absorption trails} \label{sec:ccf_phase_dependence}

\subsection{Measuring Doppler shifts through Gaussian fitting}

When combining all transits, the detection peaks of CO and H$_{\text{2}}$O appear offset with respect to the expected planet position in the $K_{\text{p}}$--$V_{\text{sys}}$ maps (Figs. \ref{fig:kp_vsys_maps}, \ref{fig:maxima_positions_svd}). This implies that the Doppler shifts of these species deviate from $v$ = \mbox{0 km/s} in the planet rest frame (e.g., \citealt{Wardenier2021,Wardenier2023}). 

To study the precise behaviour of the planet's absorption lines with orbital phase, we split the CCF signal into four quarters with bin edges $\phi = \{-13.5^\circ, -6.75^\circ, 0^\circ, 6.75^\circ, 13.5^\circ\}$. Note that we exclude the ingress and egress phases here, to make sure that the planet is fully in front of the star across each bin.

\begin{deluxetable*}{ccccccccc}
    \tablewidth{0pt} 
    \tablecaption{The phase-dependent Doppler shifts of CO and H$_{\text{2}}$O during the transit of WASP-121b, as measured in this work. The associated error $\sigma$ is computed using two different methods discussed in the text. {\color{black}{The data reported in this table are plotted in Fig. \ref{fig:data_vs_models}}.}}
    \tablehead{
        \multicolumn{1}{c}{} &
        \multicolumn{1}{c}{} &
        \multicolumn{1}{c}{\textbf{CO}} &
        \multicolumn{1}{c}{\textbf{CO}} &
        \multicolumn{1}{c}{\textbf{CO}} &
        \multicolumn{1}{c}{} &
        \multicolumn{1}{c}{\textbf{H$_{\text{2}}$O}} &
        \multicolumn{1}{c}{\textbf{H$_{\text{2}}$O}} &
        \multicolumn{1}{c}{\textbf{H$_{\text{2}}$O}} 
        \\
        \multicolumn{1}{c}{} & 
        \multicolumn{1}{c}{} &
        \multicolumn{1}{c}{Doppler shift} &
        \multicolumn{1}{c}{$\sigma_{\text{SVD}}$} &
        \multicolumn{1}{c}{$\sigma_{\text{jackknife}}$} &
        \multicolumn{1}{c}{} &
        \multicolumn{1}{c}{Doppler shift} &
        \multicolumn{1}{c}{$\sigma_{\text{SVD}}$} &
        \multicolumn{1}{c}{$\sigma_{\text{jackknife}}$} 
    }
    \label{tab:doppler_shifts}
    \startdata
        \textbf{1$^{\text{st}}$ quarter} & & $+$1.8 km/s & 1.1 km/s & 1.0 km/s & & --- & --- & --- \\
        \textbf{2$^{\text{nd}}$ quarter} & & $-$3.2 km/s & 0.8 km/s & 0.6 km/s & & $-$3.3 km/s & 1.7 km/s & 1.0 km/s \\
        \textbf{3$^{\text{rd}}$ quarter} & & $-$1.6 km/s & 0.4 km/s & 0.6 km/s & & $+$1.9 km/s & 0.8 km/s & 1.2 km/s \\
        \textbf{4$^{\text{th}}$ quarter} & & $-$3.1 km/s & 0.5 km/s & 1.5 km/s & & $+$2.3 km/s & 1.0 km/s & 1.2 km/s \\
    \enddata
\end{deluxetable*}

\begin{figure*}
\vspace{-35pt} 
\makebox[\textwidth][c]{\hspace{-25pt}\includegraphics[width=1.12\textwidth]{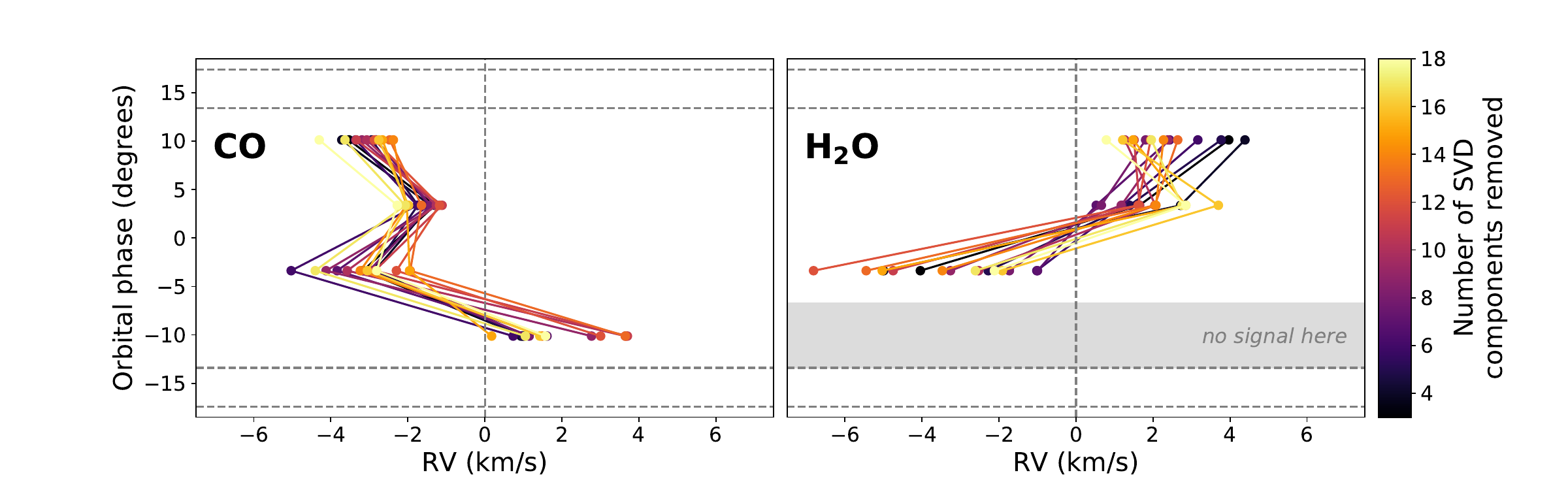}}

\vspace{-5pt}

\caption{Measured Doppler shifts for CO (left panel) and  H$_{\text{2}}$O (right panel) as a function of orbital phase angle. Different colours represent different numbers of SVD components removed from the data. We do not display error bars in this figure as the uncertainty quoted in the covariance matrix of the Gaussian fit obtained with from \texttt{scipy.optimize.curve$\_$fit} (see Fig. \ref{fig:1d_ccfs}) is much smaller than the scatter in the measurements for different numbers of SVD components. The aggregate signals with error bars can be found in Fig. \ref{fig:data_vs_models}.}
\label{fig:doppler_svd}
\vspace{10pt}
\end{figure*}

\vspace{-24pt}

Fig. \ref{fig:1d_ccfs} shows the 1D CCF signals that we obtain for CO and H$_{\text{2}}$O in the planet rest frame, for each quarter of the transit. Although the planet signals are relatively weak, each CCF (except for H$_{\text{2}}$O in the first quarter) still features a well-defined peak. As illustrated in Fig. \ref{fig:ccf_maps}, there is no H$_{\text{2}}$O signal in the first quarter of the transit, which is why we do not recover a peak in the corresponding 1D CCF.

{\color{black}{To measure the phase-dependent Doppler shifts of CO and  H$_{\text{2}}$O, we use \texttt{scipy.optimize.curve$\_$fit} to fit a Gaussian to each CCF between $\pm$10 km/s (to add more flexibility to the fit, we also allow for the baseline to be negative). This gave better results than performing the fit across a wider range of velocities. The dashed curves in Fig. \ref{fig:1d_ccfs} (right panels) show our Gaussian fits, truncated by a horizontal line at zero. The dotted vertical lines in the plots denote the peak positions inferred from the fits.}}

As mentioned previously, we perform the above analysis for removing any number of SVD components between 3 and 18. Fig. \ref{fig:doppler_svd} shows the Doppler shifts that we obtain for CO and H$_{\text{2}}$O as a function of orbital phase angle. In spite of the scatter in the values that we measure for a given species at a given phase, the qualitative behaviour of the signals is robust under the removal of different numbers of SVD components from the data, which is reassuring. To calculate the ``final'' Doppler shifts quoted in Table \ref{tab:doppler_shifts}, we take the average Doppler shift over all numbers of removed SVD components, \emph{weighted} by the detection SNR obtained with the $\sigma$-clipping method\footnote{It should be noted that the weighted average differs by \mbox{$<0.1$} km/s from the unweighted average, so both methods give essentially the same Doppler shift.} (Fig. \ref{fig:srn_svd_components}).

\subsection{Computing error bars} \label{sec:error_bars}

The uncertainties quoted in the covariance matrix of Gaussian fits to the 1D CCFs are too small ($< 0.1$ km/s) to constitute realistic error bars. Therefore, we use two alternative methods to estimate the error on the measured Doppler shifts from Table \ref{tab:doppler_shifts}. Our first error estimate, $\sigma_{\text{SVD}}$, is the square root of the weighted variance of the Doppler shifts across all numbers of removed SVD components. Again, the weights are given by the SNR values obtained from the $\sigma$-clipping method.

For the second error estimate, $\sigma_{\text{jackknife}}$, we assume that 5 SVD components are removed from the data. We then split each quarter of the transit into 5 bins, which span $\sim$1.3$^\circ$ in orbital phase\footnote{Individual frames in the dataset span $0.9^\circ$ (nights 1 and 3) and $1.1^\circ$ (night 2) in orbital phase, respectively.}. Subsequently, we measure the Doppler shift in each quarter by fitting a Gaussian to the sum of each combination of 4 bins (leaving out 1 ``sample''). Such a jackknife approach gives rise to 5 separate measurements. The error estimate is given by the standard deviation of these values.

As shown in Table \ref{tab:doppler_shifts}, both methods produce similar error estimates overall (within a factor 2). The only exception is the Doppler shift of CO in the fourth quarter of the transit (0.5 vs 1.5 km/s). Upon closer inspection, it turns out that one bin contains a large part of the signal, and when this bin is left out, the Doppler shift measurement changes by a few km/s. This is why the error obtained from the jackknife is larger.

\section{Global circulation models of WASP-121b} \label{sec:gcm_comparison}

The phase-dependent Doppler shifts of CO and H$_{\text{2}}$O observed in this work are testament to the 3D nature of UHJs, motivating the need for 3D global circulation models (GCMs) to interpret the data. To constrain the atmospheric properties of WASP-121b, we compare the measured CO and H$_{\text{2}}$O trails to four different GCM scenarios. A more quantitative retrieval study to infer chemical abundances and wind speeds (e.g., \citealt{Maguire2022,Gandhi2022,Gandhi2023,Pelletier2023,Hood2024,WeinerMansfield2024}) is left for future work ({\color{xlinkcolor}{Levens et al., in prep.}}).

\begin{figure*}
\vspace{-7pt} 
\makebox[\textwidth][c]{\hspace{-31pt}\includegraphics[width=1.12\textwidth]{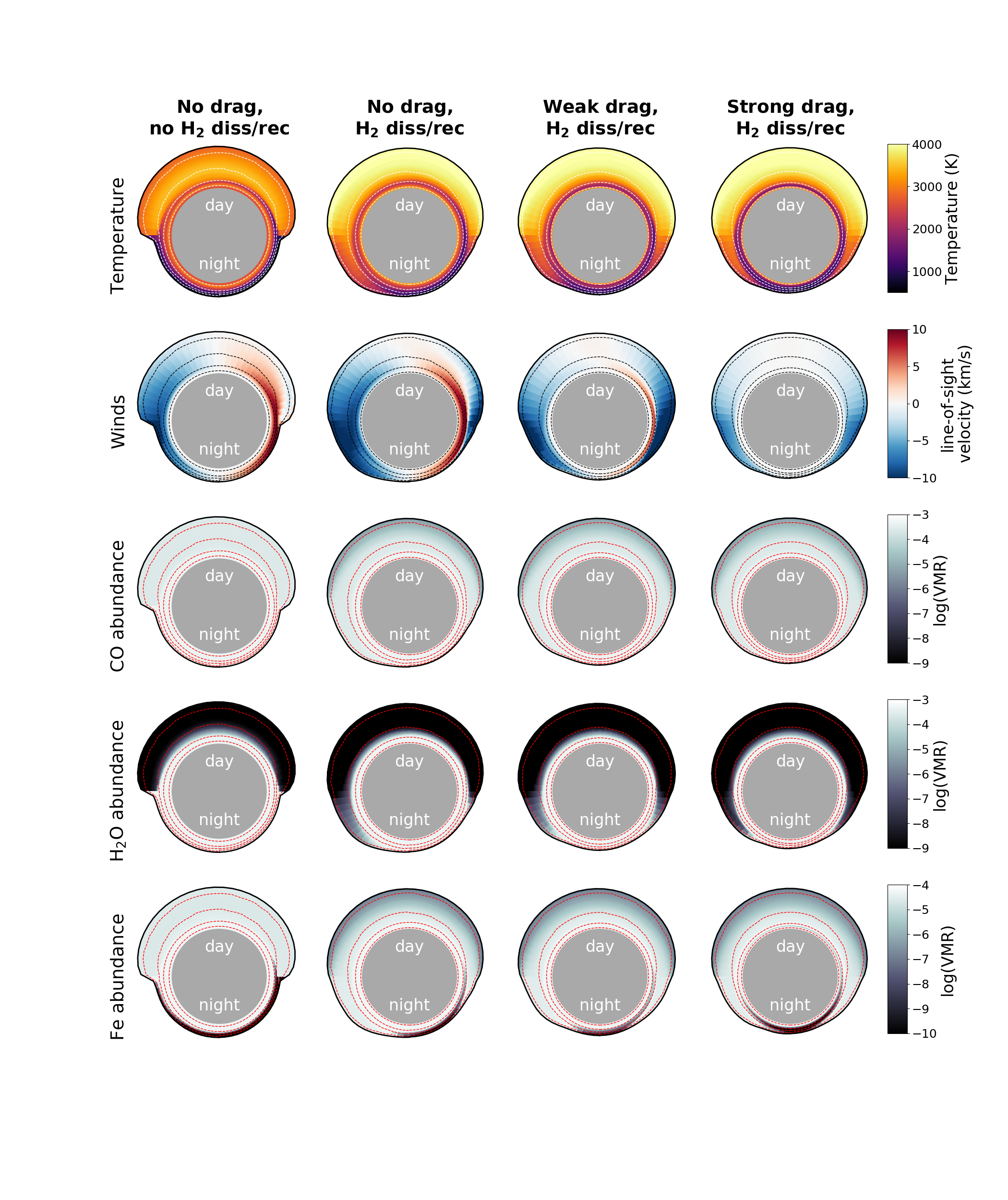}}

\vspace{-85pt}

\caption{Overview of the four GCM models of WASP-121b considered in this work. {\color{black}{The model in the first column is from \citet{Parmentier2018}, while the other three models (which account for heat transport due to hydrogen dissociation/recombination) are from \citet{Tan2024}.}} Each panel shows the equatorial plane of the planet, with the relative size of the atmosphere inflated for visualisation purposes. From top to bottom, the rows show the temperature structure, the line-of-sight velocities due to winds (at mid-transit), and the spatial distribution of CO, H$_{\text{2}}$O, and Fe, respectively. The dashed contours in each plot represent isobars with pressures $P = \{10^1$, $10^{-1}$, $10^{-3}$, $10^{-5}\}$ bar.}
\label{fig:equatorial_maps}
\end{figure*}

\subsection{Four models of WASP-121b} \label{sec:gcm_models}

We consider four SPARC/MITgcm models of WASP-121b, which are presented in Fig. \ref{fig:equatorial_maps}. {\color{black}{The SPARC/-MITgcm was first described by \citet{Showman2009}. Since, it has been widely used to study the atmospheric physics and chemistry of (ultra-)hot Jupiters (e.g., \citealt{Fortney2010,Showman2013, Kataria2013, Parmentier2018, Steinrueck2020, Tan2024}).}}

Our first model of WASP-121b is the drag-free atmosphere from \citet{Parmentier2018}. As shown in the first column in Fig. \ref{fig:equatorial_maps}, the temperature structure of this model is essentially symmetric, such that the morning and evening limbs have similar chemical compositions. The other three models are based on work by \citet{Tan2024}. In contrast to the model from \citet{Parmentier2018}, these GCMs also account for heat transport due to H$_{\text{2}}$ dissociation and recombination (e.g., \citealt{Bell2018,Komacek2018,Tan2019,Roth2021}). The idea behind this mechanism is that H$_{\text{2}}$ thermally dissociates on the dayside, after which atomic hydrogen gets advected to the nightside, where it recombines into H$_{\text{2}}$ and releases latent heat. When the atmospheric circulation is predominantly eastward, most of this heat is dumped on the evening limb, resulting in a temperature asymmetry between the eastern and western regions of the atmosphere (second column in Fig. \ref{fig:equatorial_maps}).

\begin{deluxetable}{ccccc}

{\color{black}{

    \tablewidth{0pt} 
    \tablecaption{Overview of some of the parameters of the GCMs described in Section \ref{sec:gcm_models} (see Fig. \ref{fig:equatorial_maps} for plots of the equatorial plane of each model).}
    \tablehead{
        \multicolumn{1}{c}{Parameter} &
        \multicolumn{1}{c}{Value} &
    }
     \label{tab:gcm_parameters}
      \startdata
        Orbital period & 1.1007$\times$$10^5$ s (1.274 days) \\
        Pressure range & 200 -- 2$\times$10$^{-6}$ bar \\
        Radius at bottom & 1.3038$\times$$10^8$ m (1.824 R$_{\rm Jup}$) \\
        Gravity & 8.43 m/s$^2$ \\  
        Horizontal resolution & C32 \\
        Vertical resolution & 53 layers \\
        Metallicity and C/O & 1 $\times$ solar \\
        H/H$_2$ heat transport? & $\{\times, \checkmark, \checkmark, \checkmark\}$ \\  
        Drag timescale & $\{\infty, \infty, 10^6 \ \text{s}, 10^4 \ \text{s}\}$ \\
        Radiative transfer & non-grey (see \citealt{Kataria2013}) \\
    \enddata

}}
\vspace{-15pt}
\end{deluxetable}

\vspace{-11pt}

To explore the effect of atmospheric drag (e.g., \citealt{Showman2013,Komacek2016,Parmentier2018a}), we also consider two models with drag timescales $\tau_{\text{drag}} = 10^{6}$ s (weak drag) and $\tau_{\text{drag}} = 10^{4}$ s (strong drag), which are shown in the third and fourth column of Fig. \ref{fig:equatorial_maps}. The drag timescale encapsulates a variety of physical mechanisms, such as turbulent mixing (\citealt{Li2010}), Lorentz-force braking of ionised winds in the planet's magnetic field (\citealt{Perna2010}), and Ohmic dissipation (\citealt{Perna2010a}). {\color{black}{The drag deposits energy back into the atmosphere.}} As shown in Fig. \ref{fig:equatorial_maps}, increasing the {\color{black}{drag strength}} (and thus lowering $\tau_{\text{drag}}$) slows down winds in the atmosphere\footnote{$\tau_{\text{drag}}$ is a tunable parameter in the GCM. It is included as an additional ``Rayleigh drag'' term $-\vec{v}/\tau_{\text{drag}}$ in the momentum equation solved by the dynamical core. This approach assumes that the drag force is \emph{uniform} across the whole planet atmosphere. A drag-free scenario corresponds to $\tau_{\text{drag}}\rightarrow\infty$.}. This leads to less efficient heat redistribution and a more symmetric temperature structure. Also, the equatorial jet gets suppressed -- in the strong-drag model, there is only a day-to-night flow.

Another notable difference between the model from \citet{Parmentier2018} and the models from \citet{Tan2024} is that the latter account for opacities from Fe when evaluating heating and cooling rates, making the thermal inversion on the dayside extend to much lower pressures. {\color{black}{We refer to Table 1 in \citet{Tan2024} for the full list of opacities considered in their radiative transfer and to \citet{Freedman2014} for the opacities used by \citet{Parmentier2018}.}} While the extra optical opacity causes the models with H$_{\text{2}}$ dissociation/recombination to look substantially hotter, they have nearly the same effective temperature ($\sim$2400 K) at the dayside photosphere ($\sim$0.1 bar) as the model from \citet{Parmentier2018}.

{\color{black}{Table \ref{tab:gcm_parameters} provides a summary of some other important parameters of the four SPARC/MITgcm models.  All models were run at a horizontal resolution of C32, which corresponds to roughly 128 cells in longitude and 64 cells in latitude. Before computing phase-dependent spectra of the GCMs with gCMCRT, we bin the outputs down to 32 latitudes and 64 longitudes, as was done in \citet{Wardenier2021,Wardenier2023}.}}

\subsection{Computing absorption trails}

\begin{figure*}
    \centering

    \vspace{10pt} 
    
    \begin{minipage}[c]{\textwidth}
    \makebox[\textwidth][c]{\hspace{15pt}
    \includegraphics[width=1.18\textwidth]{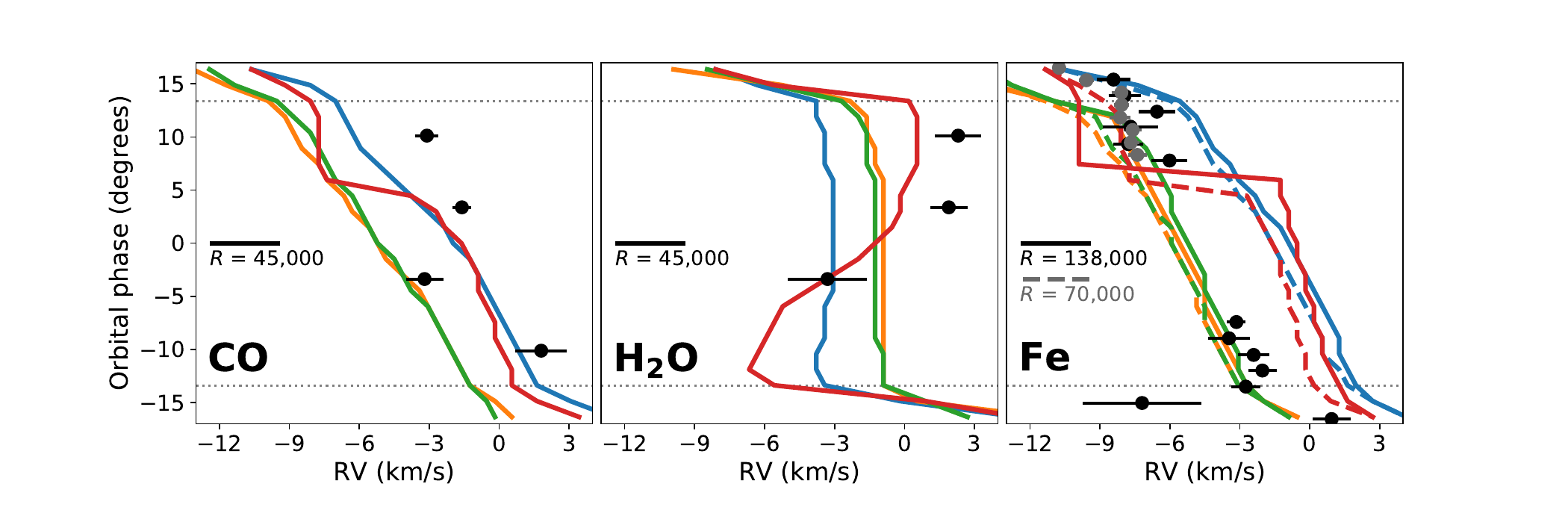}}
    \end{minipage}

     \vspace{-210pt} 

    \begin{minipage}[c]{\textwidth}
    \makebox[\textwidth][c]{
    \includegraphics[width=0.95\textwidth]{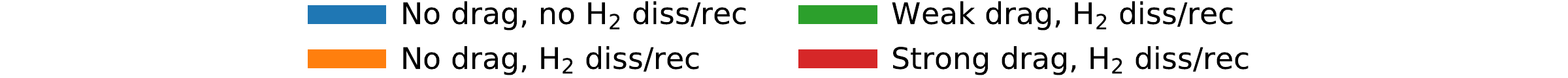}}
    \end{minipage} 

    \vspace{148pt} 

    \begin{minipage}[c]{1\textwidth}
      \vspace{25pt}
	  \caption{CCF signals of CO, H$_{\text{2}}$O, and Fe for each of the four GCM models of WASP-121b {\color{black}{(computed at 25 phases across the transit)}}. In each panel, the trails show the maximum of the CCF as a function of orbital phase. The data points in the left and middle panel depict the phase-dependent Doppler shifts of CO and H$_{\text{2}}$O measured in this work (the error bars correspond to $\sigma_{\text{SVD}}$ in Table \ref{tab:doppler_shifts}). The data points plotted in the right panel are from \citet{Borsa2021}, who observed the transit of WASP-121b with VLT/ESPRESSO in 1-UT ($R\sim$ 138,000, black points) and 4-UT ($R\sim$ 70,000, grey points) mode, respectively. The dashed and solid trails are from \emph{the same} GCM models, but represent different spectral resolutions (see also Fig. \ref{fig:full_ccf_maps_gcm}). The gap in the observed Fe trail is due to the Doppler shadow of the host star, whose photosphere ($T_{\text{eff}} \sim 6700$ K, e.g., \citealt{Daylan2021}) contains Fe but no H$_{\text{2}}$O and CO.} 
   \vspace{10pt}
   \label{fig:data_vs_models}
    \end{minipage}
\end{figure*}

Our Doppler-shift measurements of CO and H$_{\text{2}}$O with IGRINS are not the only phase-resolved transit observations of WASP-121b: \citet{Borsa2021} measured the absorption trail of Fe in the optical with VLT/ESPRESSO (see Fig. \ref{fig:data_vs_models}). We can thus rely on three species to constrain the atmospheric properties of WASP-121b.

We use gCMCRT (\citealt{Lee2022}) to compute phase-dependent transmission spectra of the four GCM models across the ESPRESSO (0.38$-$0.79 $\mu$m) and IGRINS (1.42$-$2.42 $\mu$m) bandpasses, accounting for Doppler shifts due to planet rotation and winds.  {\color{black}{All details regarding the radiative transfer and post-processing can be found in Section 2 in \citet{Wardenier2023}, so we just provide a brief summary of the gCMCRT setup below.

Before feeding the GCM outputs into gCMCRT, we map the atmospheric structures onto a 3D grid with altitude (instead of pressure) as a vertical coordinate. We account for the fact that each atmospheric column has a different scale height, which is set by local gravity, temperature, and mean-molecular weight. For each of the four WASP-121b models we simulate 25 spectra (equidistant in orbital phase) between phase angles $\pm$17.6, which also cover the ingress and egress. We assume an edge-on orbit, a semi-major axis of 0.025 AU, a stellar radius of 1.46 $R_{\odot}$, and an orbital period of 1.27 days.
At each orbital phase angle, gCMCRT simulates a transmission spectrum by randomly shooting photon packets at the planet limb and evaluating the optical depth encountered by each photon packet. In this calculation, the code accounts for Doppler shifts imparted on the opacities by the radial component of the local wind vector and planet rotation (\citealt{Wardenier2021}). The transit depth at a given wavelength is then obtained by averaging over all photon packets. We use 10$^5$ photon packets per wavelength to accurately model the shapes, depths, and shifts of the spectral lines. Because we do not explicitly treat scattering, the propagation direction of the photon packets does not change throughout the calculation. As in \citet{Wardenier2023}, the optical and infrared spectra are computed at resolutions $R = 300,000$ and $R = 135,000$, respectively. For the radiative transfer, we include the same set of continuum opacities and line species as in \citet{Wardenier2023}.}}

Before computing the CCFs, we convolve the planet spectra and the templates (see Section \ref{sec:cross-correlation}) with a Gaussian kernel corresponding to the resolution of the respective instruments. Because the ESPRESSO observations were performed both in 1-UT ($R\sim$ 138,000) and 4-UT ($R\sim$ 70,000) mode, we convolve the optical spectra and the Fe template to two resolutions, and we calculate the CCF maps for both. Fig. \ref{fig:data_vs_models} shows the CCF signals of CO, H$_{\text{2}}$O, and Fe that we obtain for each of the models, with the real data plotted on top. The full CCF maps are shown in Fig. \ref{fig:full_ccf_maps_gcm}.

\section{Discussion} \label{sec:discussion}

\subsection{The CO signal}

Indeed, as suggested by the negative $K_{\text{p}}$ offset of CO in the $K_{\text{p}}$--$V_{\text{sys}}$ map (Fig. \ref{fig:kp_vsys_maps}), the absorption lines of CO become \emph{increasingly} blueshifted during the transit of WASP-121b. This is reminiscent of the CCF signals of Fe and other refractory species that have been reported in the atmosphere of WASP-76b (\citealt{Ehrenreich2020,Kesseli2021,Pelletier2023}), whose equilibrium temperature is $\sim$250 K lower than that of WASP-121b. Our observation is also in qualitative agreement with \citet{Wardenier2023}, in which we predicted that CO should exhibit similar absorption signatures as refractory species in the atmospheres of UHJs. This is because absorption by CO mainly occurs on the dayside of the planet (see Fig. 7 in \citealt{Wardenier2023}). At the start of the transit, the CO signal is dominated by the leading (morning) limb, where the \emph{redshift} associated with planet rotation counteracts the \emph{blueshift} associated with day-to-night winds. Then, as the dayside of the trailing (evening) limb rotates into view, the Doppler shift should become more negative as planet rotation and day-to-night winds both impart a blueshift to the signal (see also Fig. 2 in \citealt{Wardenier2021}). 

As far as our GCMs of WASP-121b are concerned (Fig. \ref{fig:data_vs_models}), the drag-free and weak-drag models with H$_{\text{2}}$ dissociation/recombination produce the strongest blueshifts. The fact that both models give rise to essentially the same CCF signal indicates that weak drag has little impact on the wind speeds in the observable part of the atmosphere. Further decreasing $\tau_{\text{drag}}$ does lead to a noticeable change in the line-of-sight velocities, which is why the absorption trail of the strong-drag model is less blueshifted. The jump in the signal from $-$4 km/s to $-$7 km/s after mid-transit is driven by planet rotation (\citealt{Wardenier2023}).

When H$_{\text{2}}$ dissociation/recombination is accounted for in our models, the observed CO signal appears to be most commensurate with strong drag in the atmosphere of WASP-121b, although this model still overpredicts the observed blueshift in the second half of the transit by a few km/s. Interestingly, the drag-free model \emph{without} H$_{\text{2}}$ dissociation/recombination produces similar Doppler shifts as the strong-drag model \emph{with} H$_{\text{2}}$ dissociation/recombination. This result underscores the sensitivity of the wind profile and the resulting Doppler shifts to the treatment of heat transport in the GCM. 

Looking at Fig. \ref{fig:data_vs_models}, the Doppler-shift measurement of CO in the second quarter of the transit (just before mid-transit) appears to be somewhat of an outlier compared to the other data points. This is further corroborated by the phase-dependence of the CCF peak strength (see Fig. \ref{fig:1d_ccfs}, as well as Fig. \ref{fig:appendix_strength_fwhm} in Appendix \ref{app:A}). Throughout the transit, the signal strength is essentially constant, except in the second quarter, where the signal is a factor $\sim$2 weaker. This variation is not predicted by any of the models considered in this work (Fig. \ref{fig:full_ccf_maps_gcm}), so this data point should be interpreted with caution. 

\subsection{The H$_{\text{2}}$O signal}


Contrary to CO and Fe, the H$_{\text{2}}$O signal of WASP-121b becomes \emph{less} blueshifted during the transit -- in agreement with the positive $K_{\text{p}}$ offset in the combined $K_{\text{p}}$--$V_{\text{sys}}$ map in Fig. \ref{fig:kp_vsys_maps}. As demonstrated in the middle panel of Fig. \ref{fig:data_vs_models}, the only model that can capture the observed phase-dependence of the H$_{\text{2}}$O trail is the strong-drag model. This is not surprising -- in \citet{Wardenier2023}, we demonstrated that a decrease in blueshift can only be achieved in a scenario where planet rotation is the dominant contributor to the line-of-sight velocities (e.g., Fig. 8 in \citealt{Wardenier2023}). That is, winds must be slowed down substantially. Owing to the spatial distribution of H$_{\text{2}}$O across the atmosphere of an UHJ, most of its signal originates from the \emph{nightside}, leading to an exact opposite behaviour compared to species that are abundant on the dayside of the planet, such as CO and Fe. At the start of the transit, the trailing (evening) limb dominates the planet signal, while the contribution from the leading (morning limb) is strongest at the end of the transit.

\begin{figure*}
\vspace{-25pt} 
\makebox[\textwidth][c]{\hspace{-1pt}\includegraphics[width=1.15\textwidth]{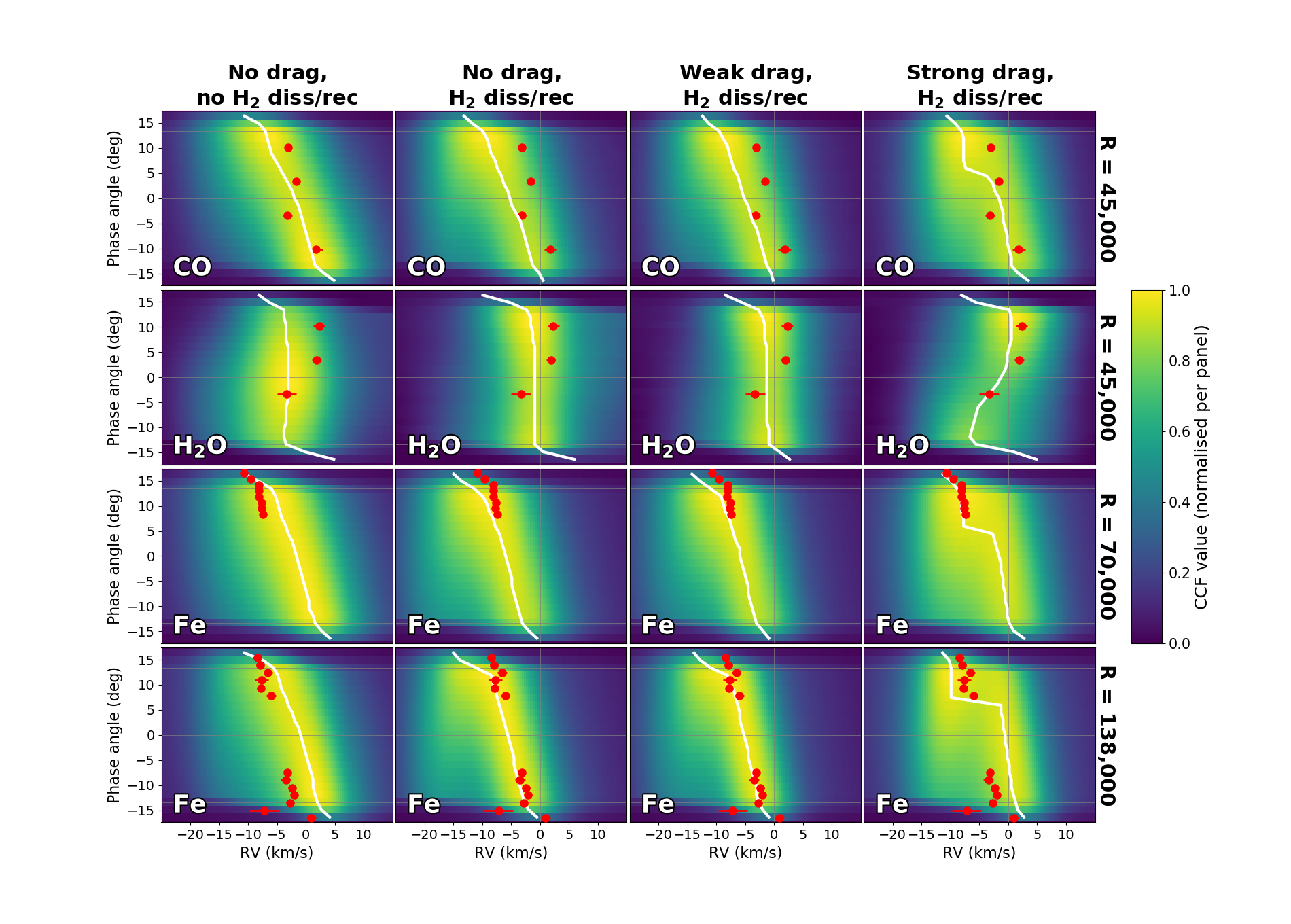}}

\vspace{-25pt}

\caption{Simulated CCF maps of CO (top row), H$_{\text{2}}$O (second row), and Fe (bottom rows, at two different spectral resolutions), based on the four GCM models considered in this work. All maps were normalised to their own maximum. The white curves in each panel indicate the maximum of the CCF and are the same as the trails plotted in Fig. \ref{fig:data_vs_models}. The red data points show the Doppler shifts measured for each species at the given spectral resolution. In the dataset from \citet{Borsa2021}, the 4-UT observations at $R\sim$ 70,000 only cover the last part of the transit, while 1-UT observations at $R\sim$ 138,000 cover the entire transit. The gap in the Fe data is due to the Doppler shadow of the star.}  
\label{fig:full_ccf_maps_gcm}
\end{figure*}

The presence of strong drag in the atmosphere of WASP-121b would be consistent with phase-curve observations from HST, TESS, and JWST in the optical and infrared (\citealt{Bourrier2020b,Daylan2021,Evans2022,Evans2023}), which revealed phase-curves offsets between zero and a few degrees eastward. 

The absence of a H$_{\text{2}}$O signal in the first quarter of the transit remains puzzling. As shown in Fig. \ref{fig:full_ccf_maps_gcm}, none of the GCM models predict zero signal at the start of the transit. However, the models accounting for H$_{\text{2}}$ dissociation/recombination do show a marginally weaker signal in the first half of the transit than in the second half, so heat transport may form part of the solution. To investigate whether noise or systematics may hamper the detection of H$_{\text{2}}$O in the first quarter (or whether the lack of signal could be of planetary origin), we perform an injection test whereby we inject a ``negative'' spectrum\footnote{We multiply each row of the data cube by $1 - (-\delta(\lambda)) = 1 + \delta(\lambda)$, with $\delta(\lambda)$ a model of the transit depth of the planet as a function of wavelength $\lambda$. We then perform a regular data analysis as described in Section \ref{sec:observations}.} into the data cube at the planet position. The spectrum is scaled such that it roughly cancels out the CCF peaks of H$_{\text{2}}$O in the other parts of the transit. This ensures that the injected signal is of the same magnitude as that of the real planet.

\begin{figure}

\vspace{12pt}

{\hspace{-13pt} \includegraphics[width=0.5\textwidth]{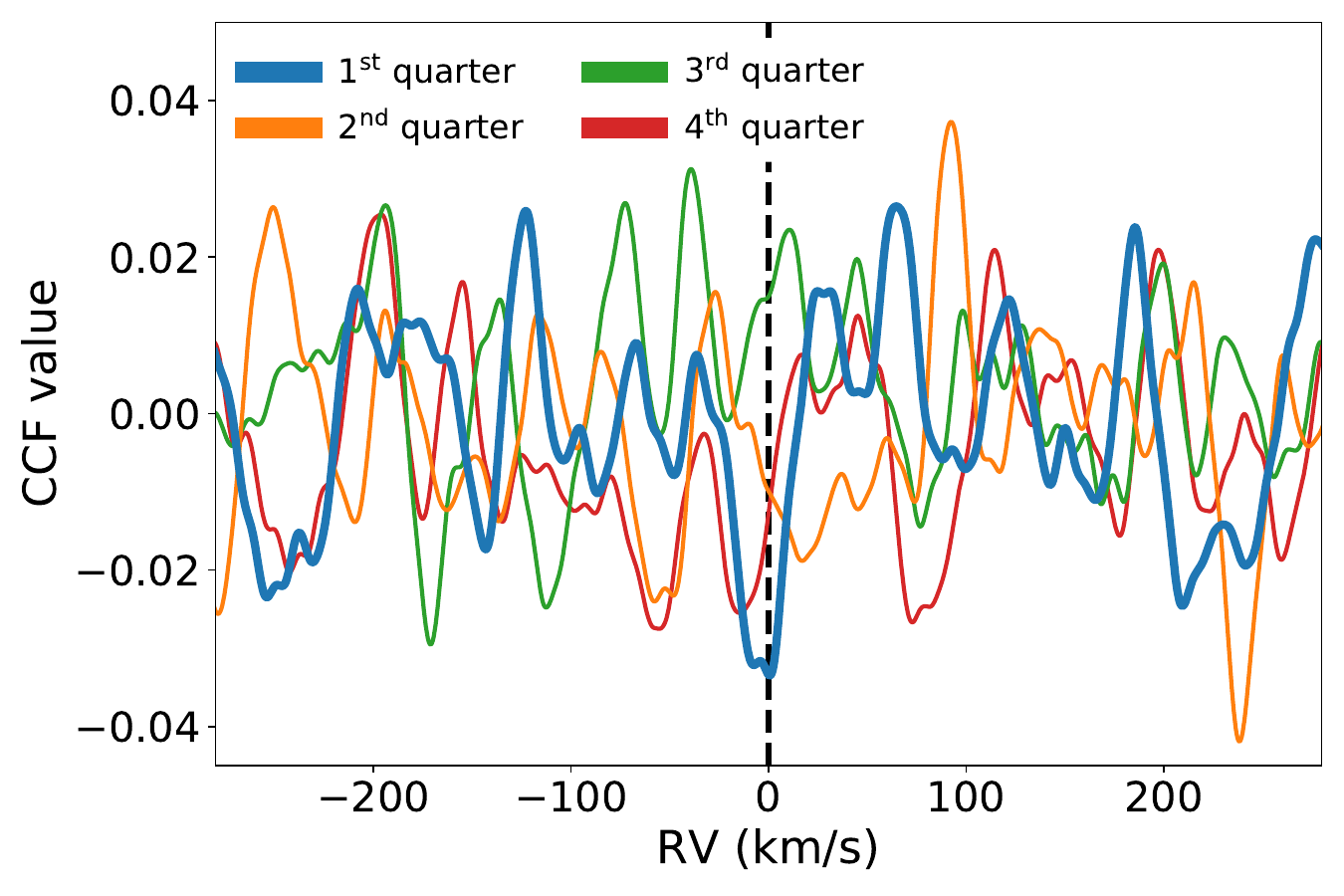}}

\vspace{-5pt}

\caption{1D CCF signal of H$_{\text{2}}$O in each quarter of the transit after the injection of a negative planet signal into the data cube (the plot is based on all three nights, with 5 SVD components removed from the data). The CCF of the first quarter features a clear minimum at 0 km/s, as opposed to the CCFs of the other quarters.}

\label{fig:water_injection}
\end{figure}

Fig. \ref{fig:water_injection} shows the results of the injection test. The CCF of the first quarter features a negative peak at the planet position. At the same time, the other parts of the transit show no clear detections as the (negative) injection cancels out the (positive) planet signal. The fact that there is a negative peak in the first quarter suggests that there is indeed little planet signal to be cancelled out. Furthermore, it demonstrates that the injected signal is detectable given the noise and the systematics of the observation. Thus, had a H$_{\text{2}}$O signal been present in the first quarter of the transit, we should have been able to measure it.

Although the injection test suggests that the lack of H$_{\text{2}}$O absorption could be physical, it cannot be fully ruled out that telluric effects play a role (see Section \ref{sec:observations}). In this light, additional observations with different barycentric-velocity offsets (that do not overlap with the first half of the planet transit) will be beneficial.

\begin{figure}

\vspace{4pt}

{\hspace{-5pt} \includegraphics[width=0.49\textwidth]{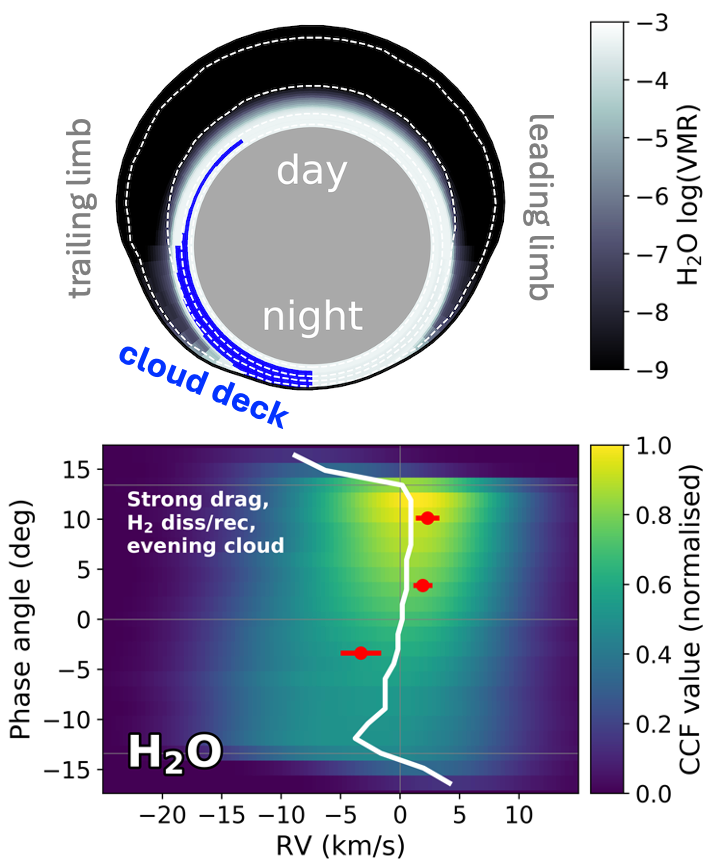}}

\vspace{-2pt}

\caption{Top: The H$_{\text{2}}$O abundance in the equatorial plane of the strong-drag model, but now with a cloud deck on the trailing (evening) limb (see Section \ref{sec:clouds} for further details). Bottom: The CCF map of H$_{\text{2}}$O obtained when including the above cloud in the strong-drag model. The signal in the first quarter of the transit is significantly muted. The red data points show our IGRINS measurements.}

\label{fig:optically_thick_cloud}
\end{figure}

\subsection{Muting the H$_{\text{2}}$O signal} \label{sec:clouds}

If the absence of a H$_{\text{2}}$O signal in the first quarter of the transit is indeed physical, it could the the result of clouds on the nightside of WASP-121b (e.g., \citealt{Helling2021}). Because CO and Fe mainly probe the dayside of the atmosphere, their signals should remain largely unaffected (\citealt{Wardenier2023}). Crucially, however, clouds should only mute the H$_{\text{2}}$O signal on the \emph{trailing} limb, but not on the leading limb. To explore this scenario, we recompute the transit spectra of the strong-drag model, but this time we include an optically thick cloud deck between longitudes 0$^\circ < \varphi <$ 180$^\circ$. Our cloud presciption is identical to that in \citet{Savel2022} and \citet{Wardenier2023}. That is, we place the cloud deck at all temperatures that lie below the Al$_2$O$_3$ condensation curve, while restricting its vertical extent to 10 scale heights (see top panel in Fig. \ref{fig:optically_thick_cloud}). We note that this is a very crude way to account for the impact of clouds, so the model should be seen as a limiting case.

The bottom panel in Fig. \ref{fig:optically_thick_cloud} shows the resulting H$_{\text{2}}$O signal. In the first quarter of the transit, H$_{\text{2}}$O absorption is indeed strongly suppressed {\color{black}{(also see Fig. \ref{fig:appendix_strength_fwhm} in Appendix \ref{app:A})}}, which is in better agreement with our observations. On the other hand, there is a larger discrepancy between the measured and modelled Doppler shifts in the second quarter of the transit. This is a consequence of the muted trailing-limb contribution. It should be noted, though, that our cloud model is by no means an quantitative fit to the data. The cloud parameters (opacity, vertical extent, etc.) could be optimized to better match the Doppler shifts, but such a retrieval is beyond the scope of this work.

Besides considering clouds, we also investigate whether chemical transport can give rise to a strong asymmetry in the H$_{\text{2}}$O abundance. The idea is that it takes a nonzero time for H and OH to recombine into H$_{\text{2}}$O as material is advected from the dayside to the nightside of the planet. Hence, the true H$_{\text{2}}$O abundances in the limb region do not just depend on pressure and temperature, but also on atmospheric circulation and recombination timescales (e.g., \citealt{Parmentier2018}). To simulate this scenario, we feed the equatorial abundances and wind speeds from the GCMs into the 2D VULCAN photochemical model (\citealt{Tsai2024}). {\color{black}{For both the weak-drag and the strong-drag model, we find that the nightside H$_{\text{2}}$O abundances only change substantially (by up to 3 orders of magnitude) in small regions of the atmosphere around 10$^{-5}$ bar (see Fig. \ref{fig:appendix_vulcan} in Appendix \ref{app:A}). Muting the H$_{\text{2}}$O signal, however, would require such changes across multiple dex in pressure (just like the cloud deck in Fig. \ref{fig:optically_thick_cloud} has a large vertical extent).}} Therefore, chemical transport alone cannot explain the muted H$_{\text{2}}$O signal in the first quarter of the transit. 


If ``evening clouds'' are indeed the most plausible mechanism behind the strongly varying  H$_{\text{2}}$O signal of WASP-121b, one would simultaneously have to explain why clouds do \emph{not} prevail on the leading (morning) limb of the planet. One scenario could be that condensates form on the trailing limb, and settle gravitationally as they are advected across the nightside. Another driver could be cloud-patchiness caused by global-scale cloud transport (e.g., \citealt{Komacek2022}). 

When it comes to confirming the cloud hypothesis, the JWST/NIRSpec phase curve of WASP-121b (\citealt{Evans2023}) will be a valuable dataset. Cloudy regions on the nightside should emit less flux than predicted by (cloud-free) GCMs, as clouds block radiation from deeper, hotter layers of the atmosphere (e.g., \citealt{Parmentier2018a,Parmentier2021,Bell2024}). The JWST dataset also includes a transit spectrum. At low resolution, evening clouds (if present) should also mute H$_{\text{2}}$O features at the start of the transit.


\subsection{The Fe signal}

The right panel in Fig. \ref{fig:data_vs_models} shows the Fe trails of our GCM models, along with the Doppler shifts of Fe measured by \citet{Borsa2021} with ESPRESSO. Especially for the strong-drag model, there is a considerable difference between the absorption trails at both ESPRESSO resolutions. This is because the trails only show the location of the \emph{maximum} CCF value as a function of phase. Fig. \ref{fig:full_ccf_maps_gcm} illustrates what is going on. At $R\sim$ 138,000, the CCF map consists of two modes, each associated with one of the planet limbs -- see \citet{Nortmann2024} for an extreme example of this effect. At such high resolution, the two modes are still ``spectrally'' separated. However, when the spectra are convolved to $R\sim$ 70,000, the modes blend into one, changing the location of the CCF maximum.

For $R\sim$ 70,000 (4-UT mode) there is impressive agreement between the ESPRESSO data and the absorption trail of the strong-drag model, especially since the GCM is by no means an optimised fit to the data. For $R\sim$ 138,000 (1-UT mode), the situation is more complicated. Firstly, the strong-drag model does not manage to reproduce the observed Doppler shifts of Fe in the first half of the transit. Rather, the data show much better agreement with the weak-drag and drag-free models that include H$_{\text{2}}$ dissociation/recombination. On the contrary, the drag-free model \emph{without} H$_{\text{2}}$ dissociation/recombination does not provide a good match -- it underpredicts the blueshift of Fe across the entire transit. In the second half of the transit, none of the GCM models produce a perfect fit to the data at $R\sim$ 138,000. While the strong-drag model matches the observed trail better in terms of shape (note the Doppler shifts during the egress phase of the transit), the weak-drag model shows better agreement in terms of overall Doppler shift.

Taken together, the Fe observations are consistent with at least some degree of drag \mbox{($10^4 \lesssim \tau_{\text{drag}} \lesssim \mbox 10^6$ s)} in the atmosphere of the planet, {\color{black}{but this drag may not necessarily be uniform}}. Moreover, as opposed to WASP-76b, the Fe trail of WASP-121b does not provide clear evidence of a strong thermochemical asymmetry between the morning and evening terminator (e.g., \citealt{Wardenier2021,Savel2022}). As illustrated in Fig. \ref{fig:equatorial_maps}, the limbs of the weak-drag and strong-drag models have very similar temperatures and compositions.

\subsection{Model limitations}

While our GCM models are able to broadly mimic the observed Doppler-shift trends for CO, H$_{\text{2}}$O, and Fe, there still exists some tension between the measured and modelled absorption signals of WASP-121b. 

One issue is that, for each of the four GCM models, the \emph{predicted} Doppler shifts of Fe and CO are very similar, within $\sim$2 km/s at every phase (see also \citealt{Wardenier2023}). However, in the first and final quarters of the transit, the \emph{observed} Doppler shifts of both species differ by 4--5 km/s. These measurements indicate that CO and Fe must (on average) probe different pressures on the dayside, and that the vertical wind shear between these layers is probably stronger than expected (e.g., \citealt{Miller-RicciKempton2012,Seidel2021}). One limitation of the GCM in this regard, is that its upper boundary lies at 2 $\mu$bar. When performing the radiative transfer at lower pressures with gCMCRT, we simply assume that the wind speeds are the same as in the upper layers of the GCM output. While the bulk of the iron lines must probe pressures $>$2 $\mu$bar (\citealt{Wardenier2021,Wardenier2023}), this could explain part of the discrepancy.

There also exists tension between the signals of H$_{\text{2}}$O and Fe. To produce the decreasing blueshift observed for H$_{\text{2}}$O, the GCM requires strong drag (i.e., low wind speeds). Even in the strong-drag scenario, our models struggle to match the $\sim$2 km/s \emph{redshift} measured in the second half of the transit. At the same time, the strong-drag model fails to reproduce the Fe signal from \citet{Borsa2021} at both VLT resolutions. The fact that the measured Doppler shifts of Fe do not change between both resolutions (see Fig. \ref{fig:data_vs_models}) favours a scenario with weaker drag.

In the light of these observations, it should be noted that our GCMs assume \emph{uniform} drag across the atmosphere of WASP-121b. In reality, however, $\tau_{\text{drag}}$ will depend on local conditions. For example, in the case of magnetic drag, $\tau_{\text{drag}}$ is a function of the local temperature, magnetic field strength, number density, and ionisation fraction (\citealt{Rauscher2013, Beltz2023,Soriano2023}). \citet{Beltz2023} showed that accounting for these dependencies with a more sophisticated magneto-hydrodynamical (MHD) prescription can have significant impact on the Doppler shifts of species measured in transmission. Thus, it may not be surprising that a model with uniform drag fails to match the signals of all species at all orbital phases, at different spectral resolutions.  

\subsection{Non-detection of OH} \label{subsec:non-detection} 

Although we find a hint of OH in the atmosphere of WASP-121b (see Figs. \ref{fig:kp_vsys_maps}, \ref{fig:appendix_kpvsys}), the associated signal (SNR $\sim$ 2.5) is not significant. This is remarkable given that OH was confidently detected (SNR $\sim$ 6.1) in the atmosphere of WASP-76b based on one transit visit with CARMENES ($R \sim 80,000$; \citealt{Landman2021}), which is on a smaller, 3.5-m telescope. Since the equilibrium temperature of WASP-121b is $\sim$250 K higher than that of WASP-76b, one explanation could be that OH has further dissociated into atomic O and H on the hotter dayside of WASP-121b, such that its features in the transmission spectrum are weaker. A retrieval would allow to shed light on the upper limits of the OH abundance in the region probed by our transit data.

Another aspect to consider is the instrument. Based on two transit visits with IGRINS, \citet{WeinerMansfield2024} also reported OH in the atmosphere of WASP-76b. However, with SNR values of $\sim$3 and $\sim$4, respectively, their detections are weaker compared to the result from \citet{Landman2021}. In this light, observations with other high-resolution spectrographs like ESO-3.6/NIRPS (\citealt{Wildi2022}) or VLT/CRIRES+ (\citealt{Dorn2023}) may help to get a better handle on the OH signal of WASP-121b.
 
\subsection{A note on atmospheric retrievals} 

Our observations of WASP-121b highlight some challenges associated with performing high-resolution \emph{transmission} retrievals in the \emph{infrared} (see also the discussion in \citealt{Wardenier2023}). In the infrared, the main absorbing species on UHJs are CO and  H$_{\text{2}}$O, which have distinct 3D spatial distributions across the atmosphere. Therefore, the absorption lines of both species will probe different temperatures (CO is more sensitive to the dayside, while H$_{\text{2}}$O is more sensitive to the nightside), but they will also be subject to different phase-dependent Doppler shifts. This leads to species-dependent peak offsets in the $K_{\text{p}}$--$V_{\text{sys}}$ map (e.g., Fig. \ref{fig:kp_vsys_maps}).

The fact that CO and H$_{\text{2}}$O probe different parts of the atmosphere suggests that a single temperature (profile) and a single pair of ($\Delta K_{\text{p}}$, $\Delta V_{\text{sys}}$) parameters may not be sufficient to accurately represent the physics underlying the observation in a forward model. To our knowledge, it has not been studied how abundance measurements from high-resolution transmission retrievals are affected when assuming the same Doppler shift and/or temperature for CO and H$_{\text{2}}$O.

{\color{white}{---}}

\section{Summary and conclusion} \label{sec:conclusion}

In this work, we presented three transit observations of the ultra-hot Jupiter WASP-121b with \mbox{GEMINI-S/}IGRINS. We demonstrated that the instrument is capable of resolving the absorption signals of CO and H$_{\text{2}}$O with orbital phase. To our knowledge, these measurements are the first of their kind in the infrared. To interpret the absorption trails of CO and H$_{\text{2}}$O, as well as the Doppler shifts of Fe previously measured by \citet{Borsa2021} with VLT/ESPRESSO, we compared the data to simulated cross-correlation signals based on the outputs of global circulation models (GCMs).

{\color{black}{Our observations show that phase-resolved transmission spectroscopy is a powerful technique to characterize the 3D nature of (ultra-)hot Jupiters, especially in tandem with theoretical predictions from GCMs. Owing to the unique spatial distribution of different chemical species across the planet, different absorption trails provide complementary information about the 3D structure and dynamics of the atmosphere. Instruments on the next generation of ground-based telescopes, such as the \mbox{E-ELT}, will be able to resolve the absorption trails of transiting gas giants with a much better precision and phase resolution, so observations like these will become the standard in future high-resolution spectroscopy studies.}}  

Our main findings are summarised below: 

\begin{enumerate}

\item[$\bullet$] CO and H$_{\text{2}}$O are subject to different phase-dependent Doppler shifts in the atmosphere of WASP-121b. CO absorption lines become \emph{more} blueshifted during the transit (similar to refractories such as Fe), while H$_{\text{2}}$O lines become \emph{less} blueshifted, and even show a redshift in the second half of the observation. These qualitative trends, driven by a combination of planet rotation and the 3D distribution of a species across the atmosphere, are in agreement with previous modelling work from \citet{Wardenier2023} and are robust under the removal of different numbers of SVD components from the data.

\item[$\bullet$] While the CO signal strength is roughly constant with orbital phase, there is no H$_{\text{2}}$O signal in the first quarter of the transit. None of our (cloud-free) GCM models predict such strongly muted H$_{\text{2}}$O absorption at the start of the observation. The lack of H$_{\text{2}}$O signal could be due to systematics/tellurics in the data (warranting extra observations), but it can also be explained by a cloud on the evening limb of WASP-121b. 


\item[$\bullet$] The absorption trails of CO, H$_{\text{2}}$O, and Fe are all consistent with the presence of drag in the atmosphere of WASP-121b. However, none of our four models produce a perfect match to all the data simultaneously, potentially hinting at missing physics in the GCM (such as spatially dependent magneto-hydrodynamics, e.g., \citealt{Beltz2023}). The presence of drag in the atmosphere of WASP-121b is in agreement with recent phase-curve observations of the planet, which revealed relatively small hotspot offsets. Also, the Fe signal does not suggest a strong thermochemical asymmetry between the morning and evening limb of WASP-121b, as opposed to WASP-76b. 

\item[$\bullet$] CO is unambiguously detected in each of the individual IGRINS visits. For H$_{\text{2}}$O, we only recover a clear signal when adding the three observations together. We do not find significant evidence for other chemical species (including OH) in the atmosphere of WASP-121b with our data.

\end{enumerate}

\begin{acknowledgments}
J.P.W. acknowledges support from the Trottier Family Foundation via the Trottier Postdoctoral Fellowship, the Wolfson-Harrison UK Research Council Physics Scholarship, and the Science and Technology Facilities Council (STFC). M.R.L. and J.L. Bean acknowledge support from NASA XRP grant 80NSSC19K0293 and NSF grant AST-2307177. M.W.M. acknowledges support from NASA Hubble Fellowship grant HST-HF2-51485.001-A. J.L. Birkby acknowledges funding from the European Research Council (ERC) under the European Union’s Horizon 2020 research and innovation program under grant agreement no. 805445. E.K.H. Lee is supported by the SNSF Ambizione Fellowship grant (\#193448). We thank David Ehrenreich, Ray Pierrehumbert, Romain Allart, and Antoine Darveau-Bernier for helpful discussions, and Francesco Borsa for kindly sharing the VLT/ESPRESSO measurements shown in Figs. \ref{fig:data_vs_models}, \ref{fig:full_ccf_maps_gcm}, and \ref{fig:appendix_strength_fwhm}. We are grateful to Greg Mace for serving as the IGRINS point of contact. This work made use of computing resources at the AOPP subdepartment at the University of Oxford. We thank the anonymous referee for thoughtful comments that helped improve the quality of the manuscript. 
\end{acknowledgments}

%

\vspace{5mm}
\facilities{IGRINS/GEMINI-S (\citealt{Yuk2010,Mace2018}), VLT/ESPRESSO (\citealt{Pepe2010})}


\software{\texttt{matplotlib} (\citealt{Hunter2007}), \texttt{numpy} (\citealt{Harris2020}), \texttt{scipy} (\citealt{Virtanen2020}), \texttt{IGRINS$\_$transit} (\citealt{igrins_transit}), gCMCRT (\citealt{Lee2022}), SPARC/MITgcm (\citealt{Showman2009}), VULCAN (\citealt{Tsai2024})}




\appendix

\section{Supplementary figures}
\label{app:A}

\begin{figure*}
\vspace{-5pt} 
\makebox[\textwidth][c]{\hspace{19pt}\includegraphics[width=1.17\textwidth]{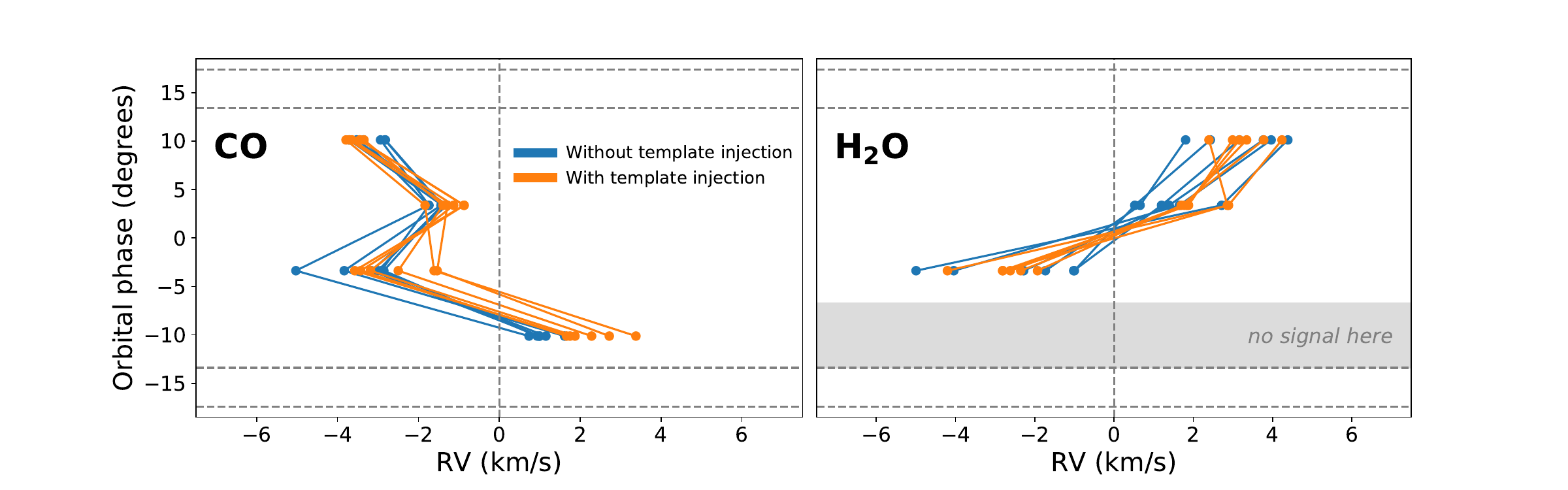}}

\vspace{-8pt}

\caption{Results of the template injection test described in Section \ref{sec:ccf_maps}. The plots show the phase-dependent Doppler shifts that we measure through our regular analysis (in blue) versus injecting the template into the main SVD components before performing the cross-correlation (in orange). For both approaches, each trail pertains to a different number of SVD components $n$, with  $3 \leq n \leq 8$.}
\vspace{5pt}
\label{fig:appendix_doppler_comparison}
\end{figure*}

In this appendix, we present four supplementary figures. \textbf{Figure \ref{fig:appendix_doppler_comparison}} shows the results of the template injection test described in Section \ref{sec:ccf_maps}. Rather than cross-correlating the data with the template spectrum directly, we first inject the template into the main SVD components of the observation (containing the bulk of the stellar/telluric signals), and then recover it through a second SVD. Both for CO and H$_{\text{2}}$O, we find that the injected and the non-injected templates give rise to very similar Doppler-shift trends, showing that the planet signals are robust across both analyses. 

\begin{figure*}
\vspace{-5pt} 
\makebox[\textwidth][c]{\hspace{19pt}\includegraphics[width=1.17\textwidth]{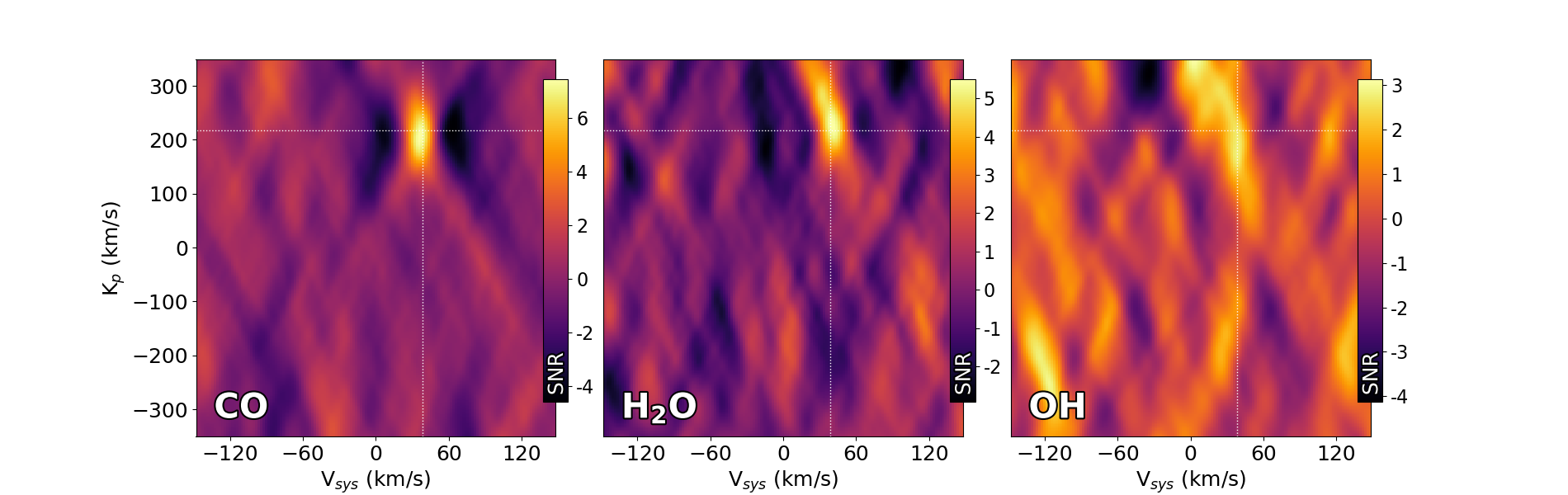}}

\vspace{-5pt}

\caption{Same plots as in the right panel of Fig. \ref{fig:kp_vsys_maps} (data from all nights combined), but now on a much larger domain. CO and H$_{\text{2}}$O show up as clear detections near the expected planet position ($K_{\text{p}}$ $\sim$ 218 km/s, $V_{\text{sys}}$ $\sim$ 38 km/s). Cross-correlation with an OH template produces a faint signal near the expected planet position, but it is not significant.}

\vspace{20pt}

\label{fig:appendix_kpvsys}
\end{figure*}

\textbf{Figure \ref{fig:appendix_kpvsys}} shows the $K_{\text{p}}$--$V_{\text{sys}}$ maps of CO, H$_{\text{2}}$O, and OH, plotted on a larger domain than in Fig. \ref{fig:kp_vsys_maps}. Our detections of CO and H$_{\text{2}}$O are significant (SNR $\gtrsim$ 5), while we only find a tentative hint of OH.

\begin{figure*}
\vspace{-20pt} 
\makebox[\textwidth][c]{\hspace{19pt}\includegraphics[width=0.7\textwidth]{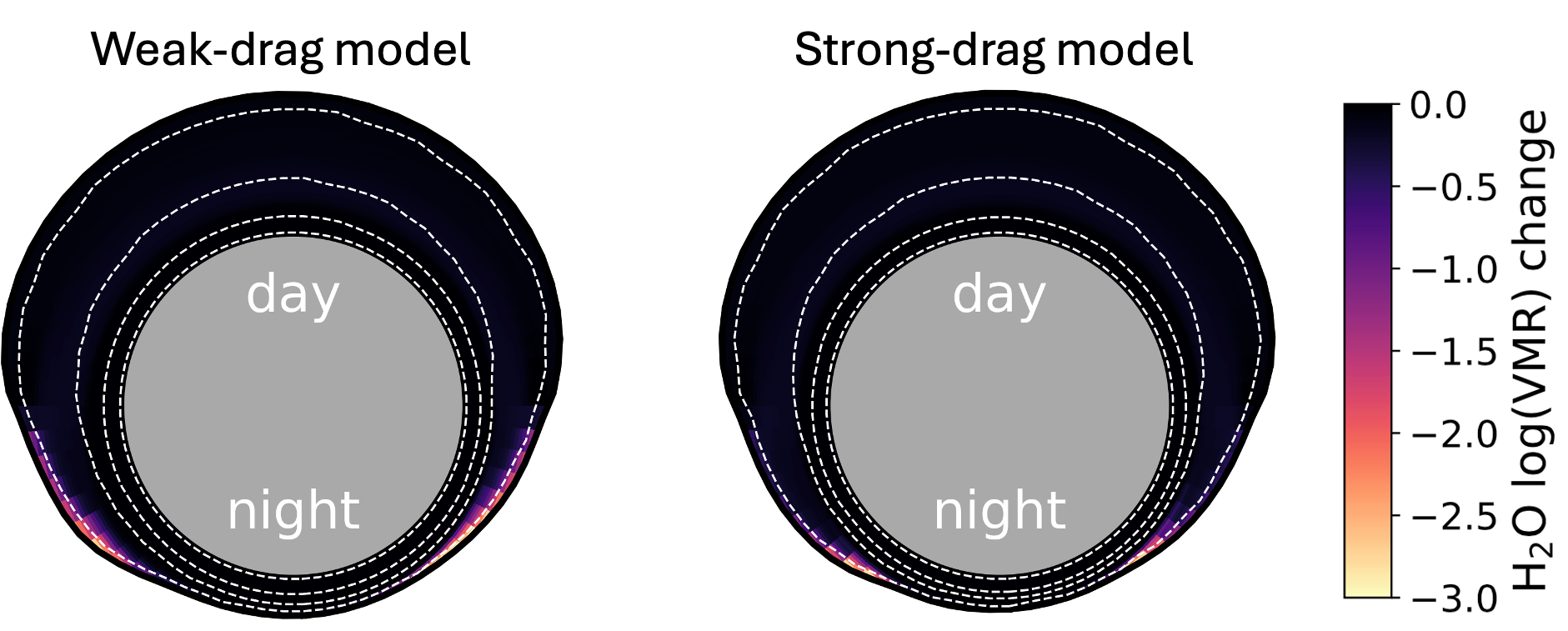}}

\vspace{0pt}

\caption{{\color{black}{Equatorial-plane plots summarising the outcomes of the 2D VULCAN (\citealt{Tsai2024}) simulations discussed in Section \ref{sec:clouds}. The plots show the change in H$_{\text{2}}$O abundance when going from a model with equilibrium chemistry to a model with disequilibrium chemistry (simulated with VULCAN). The left panel pertains to our weak-drag GCM and the right panel to the strong-drag GCM.}}}

\vspace{20pt}

\label{fig:appendix_vulcan}
\end{figure*}

{\color{black}{\textbf{Figure \ref{fig:appendix_vulcan}} shows the results of the 2D VULCAN simulation described in Section \ref{sec:clouds}. As demonstrated in the plot, chemical kinetics only has a strong impact on the water abundances in small regions around 10$^{-5}$ bar, which is not enough to substantially impact the H$_{\text{2}}$O signal of the planet.}}   

\begin{figure*}
    \centering

    \vspace{-15pt} 
    
    \begin{minipage}[c]{\textwidth}
    \makebox[\textwidth][c]{\hspace{15pt}
    \includegraphics[width=1.18\textwidth]{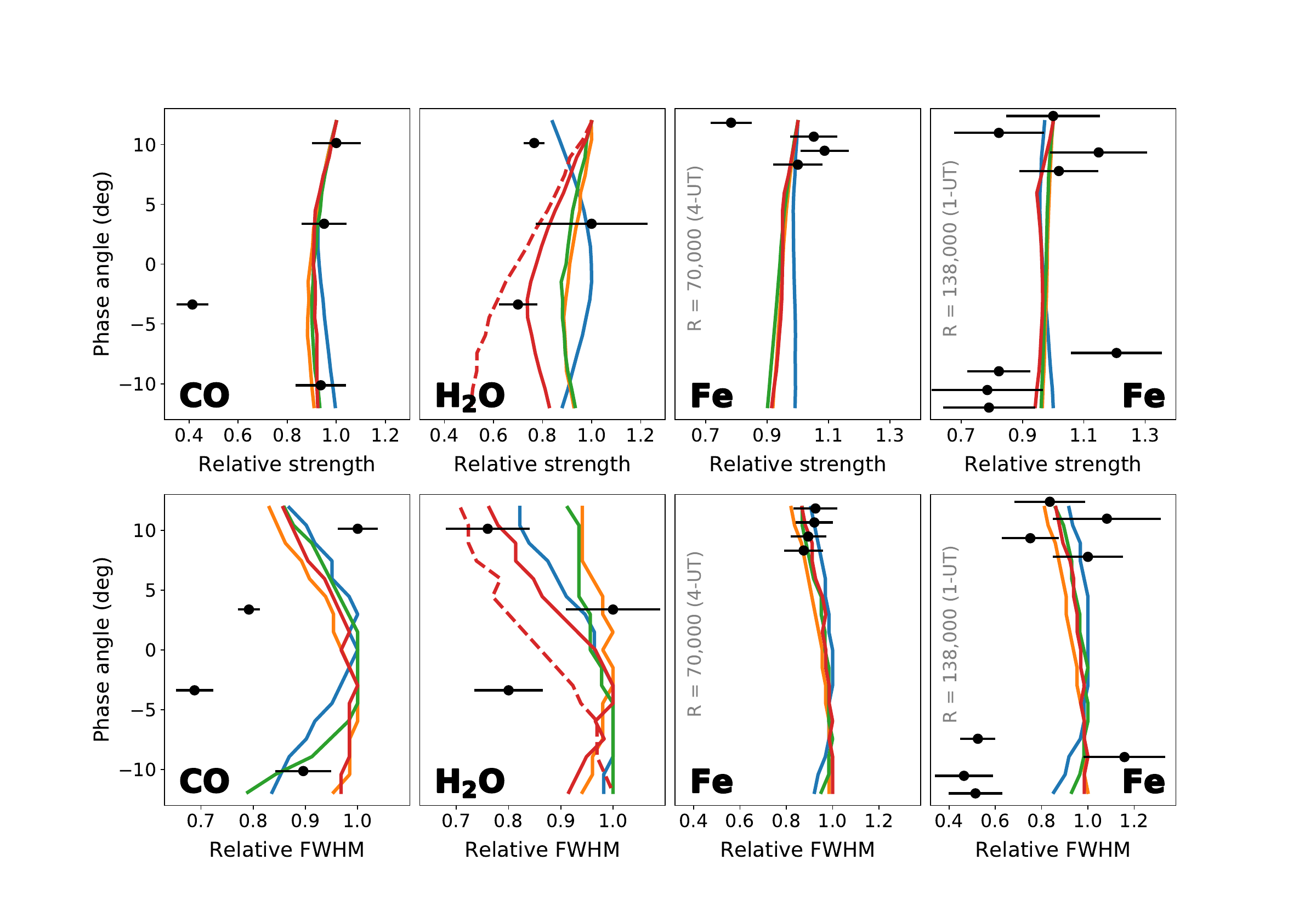}}
    \end{minipage}

     \vspace{-400pt} 

    \begin{minipage}[c]{\textwidth}
    \makebox[\textwidth][c]{
    \includegraphics[width=0.95\textwidth]{figures/data_vs_model_legend.pdf}}
    \end{minipage} 

    \vspace{338pt} 

    \begin{minipage}[c]{1\textwidth}
      \vspace{25pt}
	  \caption{Top: Comparison between the phase-dependent signal strengths observed for CO, H$_{\text{2}}$O, and Fe, and the signal strengths of the GCM models presented in Fig. \ref{fig:full_ccf_maps_gcm} (see legend). The data points for CO and  H$_{\text{2}}$O are from this work (5 SVD components removed), while the data points for Fe are from \citet{Borsa2021}, who observed WASP-121b with VLT/ESPRESSO at two different spectral resolutions $R$. All signal strengths are \emph{relative} -- the model strengths have a maximum value of unity, while the data points and their error bars were scaled such that they roughly overlap with the models. The red dashed line in the second panel shows the signal strength of the strong-drag model with evening clouds (see also Fig. \ref{fig:optically_thick_cloud}). Bottom: Similar plots showing the phase-dependent FWHM of the CCF peak for the data and the GCM models.} 
   \vspace{10pt}
   \label{fig:appendix_strength_fwhm}
    \end{minipage}
\end{figure*}

\textbf{Figure \ref{fig:appendix_strength_fwhm}} shows a further comparison between our IGRINS data (for CO and H$_{\text{2}}$O) and the VLT/ESPRESSO data from \citet{Borsa2021} (for Fe) in terms of signal strength and FWHM of the CCF peak. Performing an absolute comparison between the data and the models is tricky, mainly because the models do not account for observational noise. Also, because we cross-correlate the data with template spectra (Section \ref{sec:ccf_maps}) instead of a binary mask (e.g, \citealt{Ehrenreich2020,Borsa2021}), the strength and FWHM of the resulting CCF are not the same as the (average) strength and FWHM of the absorption lines in the spectrum, so their values are somewhat arbitrary. To circumvent these issues but still gain insight into the phase-dependent behaviour of the data and the models, we plot \emph{relative} values instead (see caption of Fig. \ref{fig:appendix_strength_fwhm} for further details). To obtain the data points for CO and H$_{\text{2}}$O in Fig. \ref{fig:appendix_strength_fwhm}, we removed 5 SVD components from the data. The error bars were obtained from a jackknife, as described in Section \ref{sec:error_bars}. 


\bibliography{citations}{}
\bibliographystyle{aasjournal}



\end{document}